\documentclass[
 aps,prb,
 twocolumn,
 superscriptaddress,
 longbibliography,
 nofootinbib,
 floatfix
]{revtex4-2}

\usepackage{amsmath,amssymb,amsfonts}
\usepackage{booktabs}
\usepackage{xcolor}
\usepackage{graphicx}
\usepackage{hyperref}
\hypersetup{colorlinks=true,linkcolor=blue,citecolor=blue,urlcolor=blue}
\usepackage{orcidlink}

\newcommand{\Jp}{J_\perp}
\newcommand{\rp}{r_\perp}
\newcommand{\qs}{q^\star}
\newcommand{\up}{\uparrow}
\newcommand{\dn}{\downarrow}
\newcommand{\authornote}{Contact:
\href{mailto:nttien@ctu.edu.vn}{nttien@ctu.edu.vn}}

\begin{document}

\title{Onset of a quantum Fisher--Selke sequence in the transverse-field ANNNI model and its bilayer}

\author{Tuan-Vu Truong\,\orcidlink{0009-0005-6357-6060}}
\affiliation{College of Natural Sciences, Can Tho University, 3-2 Road, Can Tho City
94000, Vietnam}
\affiliation{Identity Quantum Computing JSC, 117 Tran Duy Hung Street, Hanoi 10000,
Vietnam}

\author{Vu-Quoc-Minh Nguyen\,\orcidlink{0009-0007-2748-5107}}
\affiliation{Faculty of Electronics and Telecommunications, University of Science,
Vietnam National University Ho Chi Minh City (VNU-HCM), Ho Chi Minh City 70000,
Vietnam}
\affiliation{Identity Quantum Computing JSC, 117 Tran Duy Hung Street, Hanoi 10000,
Vietnam}

\author{Hoang-Long Nguyen\,\orcidlink{0009-0005-7922-9266}}
\affiliation{Identity Quantum Computing JSC, 117 Tran Duy Hung Street, Hanoi 10000,
Vietnam}

\author{Thanh-Tien Nguyen\,\orcidlink{0000-0002-6678-3787}}
\thanks{\authornote}
\affiliation{College of Natural Sciences, Can Tho University, 3-2 Road, Can Tho City
94000, Vietnam}

\date{\today}

\begin{abstract}
We study the transverse-field axial next-nearest-neighbor Ising (ANNNI) model on the square lattice
and on a bilayer. At the classical multiphase point $\kappa=J_2/J_0=1/2$, every configuration of straight domain walls
that separates domains of at least two sites is a ground state. The walls are rigid, so perturbation theory in the
transverse field $g=\Gamma/J_0$ can be carried out for each configuration separately, and we take it to order $g^8$.
Quantum fluctuations select $\langle3\rangle$ ($\up\up\up\dn\dn\dn$) at order $g^2$, because spins next to a wall gain the
most energy from virtual flips. The structures $\langle23\rangle$, $\langle223\rangle$ and $\langle2223\rangle$ then
open between $\langle3\rangle$ and the antiphase $\langle2\rangle$ at orders $g^4$, $g^6$ and $g^8$, in the order of
the Fisher--Selke sequence of the classical three-dimensional model. A second layer suppresses each successive phase more
strongly, beyond the stiffening of the local fields. We test the two widest phases with stochastic series expansion
quantum Monte Carlo, comparing the free energies of the competing ordered states by thermodynamic integration, since
annealing freezes in the modulation that forms at the ordering transition. The simulations reproduce the
$\langle3\rangle$ wedge to about $1\%$, and in a pre-registered test $\langle23\rangle$ is stable at six of seven points,
with the seventh resolved by post-hoc runs.
\end{abstract}

\maketitle

\section{Introduction}
\label{sec:intro}

The axial next-nearest-neighbor Ising (ANNNI) model is the simplest lattice model in which
competing ferromagnetic and antiferromagnetic interactions produce spatially modulated
order~\cite{elliott1961phenomenological,selke1988annni}. Along one axis, a ferromagnetic
nearest-neighbor coupling $J_1$ competes with an antiferromagnetic next-nearest-neighbor coupling
$J_2$. At zero temperature the ferromagnet is stable for $\kappa=J_2/J_1<1/2$ and the antiphase
$\langle2\rangle$, $\up\up\dn\dn$, for $\kappa>1/2$, and at the multiphase point $\kappa=1/2$ a large
number of structures are degenerate. In the three-dimensional classical model, thermal fluctuations
lift this degeneracy and generate an infinite sequence of commensurate phases that spring from the
multiphase point~\cite{fisher1980infinitely,bak1980commensurate,selke1988annni}. In two dimensions
the modulated region is instead dominated by a floating incommensurate phase with algebraic
correlations~\cite{villain1981two,pokrovsky1979ground}, whose extent was settled only
recently~\cite{shirakura2014,hu2021resolving,keiser2026}.

A transverse field $\Gamma$ adds quantum fluctuations. The transverse-field ANNNI chain is related to
the classical two-dimensional model by the Suzuki--Trotter mapping~\cite{suzuki1976relationship}
and has been studied in great
detail~\cite{rieger1996one,chandra2007floating,beccaria2007floating,suzuki2013quantum}; the width of
its floating phase was long disputed~\cite{chandra2007floating,beccaria2007floating}. In two and
more dimensions the quantum ANNNI model has been treated by mean-field and random-phase approximations
in three dimensions~\cite{sen1989ising}, by a path-integral estimate of the ferromagnetic boundary on the
square and cubic lattices for $\kappa<1/2$~\cite{sen1995order}, by large-spin perturbation theory for an
anisotropic Heisenberg analog~\cite{harris1995quantum,harris1995quantumb}, and, for quantum rotors, by field
theory of the quantum Lifshitz point~\cite{dutta1997quantum}. To our knowledge there is no unbiased numerical study of the
transverse-field ANNNI model with axial frustration on the square lattice or on its bilayer
extension. Related frustrated transverse-field Ising models with diagonal couplings have been
studied with quantum Monte
Carlo~\cite{iglovikov2013,kato2015quantum,biswas2016quantum} and, for bilayers, with cluster
mean-field theory~\cite{kellermann2019quantum,roos2024frustrated,roos2026persistent}; frustrated
transverse-field Ising models are also realized on programmable quantum
annealers~\cite{king2018observation}.

Here we consider two square layers with axial frustration, coupled by an interlayer exchange, and
ask two questions. What selects the ordered state near the multiphase point once quantum
fluctuations are present, and how does the second layer change the answer? At the multiphase point
of the square-lattice model, the in-plane ferromagnetic coupling perpendicular to the frustrated
axis forces domain walls to be straight lines, so the degeneracy is subextensive and walls cannot
move at low order in $\Gamma$. This makes the low-field problem accessible to perturbation theory.
At second order in $\Gamma$ we find that quantum fluctuations select the $\langle3\rangle$ structure
$\up\up\up\dn\dn\dn$ in a wedge around $\kappa=1/2$ whose width grows as $\Gamma^2$, and that the
interlayer coupling narrows the wedge. At fourth order the structure $\langle23\rangle$ opens on the
antiphase side of the wedge, at sixth order $\langle223\rangle$, and at eighth order $\langle2223\rangle$.
This is the order of appearance in the low-temperature analysis of the classical three-dimensional model by
Fisher and Selke~\cite{fisher1980infinitely,fisher1981low}. We establish these four steps and make no claim
about how the sequence continues. The result differs both from the one-dimensional chain, where
walls are mobile and the degeneracy gives way to a floating phase~\cite{rieger1996one}, and from the
large-spin Heisenberg analog, where the phases $\langle3\rangle,\langle4\rangle,\langle5\rangle,\dots$ appear
between the antiphase and the ferromagnet~\cite{harris1995quantum,harris1995quantumb}.

We test these predictions with stochastic series expansion (SSE) quantum Monte
Carlo~\cite{sandvik1999stochastic,sandvik2003stochastic,sandvik2010computational} at $T=0.15\,J_0$.
Frustration makes the modulated region hard to equilibrate: cluster updates percolate at small
field, and annealing from the paramagnet selects the modulation that forms at the ordering
transition and cannot relax to a different commensurate structure at lower field. We therefore
follow each competing ordered state as a separate branch and compare the branches by their free
energies, obtained by thermodynamic integration in the field.

Up to $g/(2+r)\approx0.5$ the free energies confirm the $\langle3\rangle$ wedge: its slope agrees with the prediction
to about $1\%$ for both $r$. In a test whose design
and analysis were fixed before the simulations, $\langle23\rangle$ lies below both $\langle2\rangle$ and
$\langle3\rangle$ at six of seven points. The seventh, with the smallest predicted margin, misses the
pre-registered criterion; post-hoc runs with more statistics resolve $\langle23\rangle$ there as well. On a common
lattice, at four points and with replicated runs in the bilayer, both edges of the window and both free-energy gains
agree with the eighth-order series within their statistical errors. On the $20\times30$ lattices of the test, at the standard number of sweeps, the bilayer gain of $\langle23\rangle$ over
$\langle3\rangle$ comes out about $10^{-4}J_0$ per spin larger; with four times the sweeps it agrees with the common
lattice and with the series. The phases
$\langle223\rangle$ and $\langle2223\rangle$ are too narrow to resolve numerically.

Section~\ref{sec:model} defines the model and its symmetries. Section~\ref{sec:pt} derives the
low-field phase diagram near the multiphase point. Section~\ref{sec:methods} describes the SSE
simulations, the branch protocol and their validation, and Sec.~\ref{sec:results} presents the
results. Section~\ref{sec:conclusion} concludes.

\section{Model}
\label{sec:model}

\subsection{Hamiltonian}

\begin{figure*}
\includegraphics[width=\textwidth]{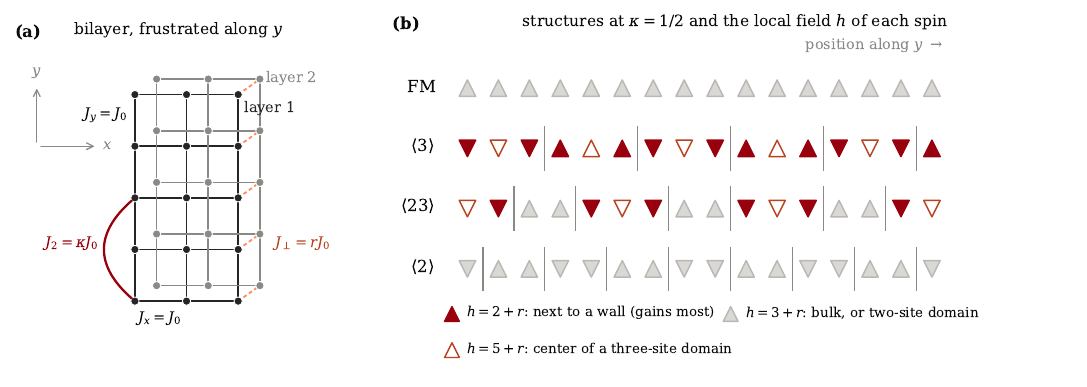}
\caption{(a) The bilayer of Eq.~\eqref{eq:H}. Each layer is a square lattice with ferromagnetic couplings
$J_x=J_y=J_0$ and an antiferromagnetic coupling $J_2=\kappa J_0$ between next-nearest neighbors along the
frustrated axis $y$ (one such bond is drawn); the layers are coupled by $J_\perp=rJ_0$. (b) Straight-wall
structures along $y$ at $\kappa=1/2$: the ferromagnet, $\langle3\rangle$, $\langle23\rangle$ and the antiphase
$\langle2\rangle$. Triangles give the spin direction and vertical lines the domain walls. Each spin is shaded
by its local field $h$ of Eq.~\eqref{eq:h} (Table~\ref{tab:h}). Spins next to a wall ($h=2+r$) gain the most
energy from virtual flips, spins in the bulk or in a two-site domain have $h=3+r$, and the center of a
three-site domain has $h=5+r$.}
\label{fig:model}
\end{figure*}

We consider two square layers $\ell=1,2$ of $L_x\times L_y$ sites $(i,j)$ with periodic boundaries
and $N=2L_xL_y$ spins,
\begin{align}
\hat H={}&-J_x\sum_{i,j,\ell}\hat\sigma^z_{i,j,\ell}\hat\sigma^z_{i+1,j,\ell}
         -J_y\sum_{i,j,\ell}\hat\sigma^z_{i,j,\ell}\hat\sigma^z_{i,j+1,\ell}
         \nonumber\\
        &+J_2\sum_{i,j,\ell}\hat\sigma^z_{i,j,\ell}\hat\sigma^z_{i,j+2,\ell}
         -\Jp\sum_{i,j}\hat\sigma^z_{i,j,1}\hat\sigma^z_{i,j,2}
         \nonumber\\
        &-\Gamma\sum_{i,j,\ell}\hat\sigma^x_{i,j,\ell}.
\label{eq:H}
\end{align}
The frustrated axis is $y$ [Fig.~\ref{fig:model}(a)]. We set $J_x=J_y=J_0>0$ and use
\begin{equation}
\kappa=\frac{J_2}{J_0},\qquad g=\frac{\Gamma}{J_0},\qquad \rp=\frac{\Jp}{J_0}.
\label{eq:params}
\end{equation}
The single-layer model is the case $\rp=0$, in which the two layers decouple. The off-diagonal
matrix elements $-\Gamma$ are nonpositive, so the model is stoquastic and has no sign problem in
quantum Monte Carlo.

\subsection{Symmetries}
\label{sec:symmetries}

Two unitary transformations relate different couplings exactly. The layer flip
$\hat U_2=\prod_{i,j}\hat\sigma^x_{i,j,2}$ reverses $\hat\sigma^z$ in layer~2, changes the sign of
the interlayer term and leaves all other terms invariant, so $\hat H(\kappa,g,\rp)$ and
$\hat H(\kappa,g,-\rp)$ have the same spectrum and the same in-plane correlations; only the
stacking of the two layers is reversed. The alternate-row spin flip
$\hat V=\prod_{i,\,j\ \mathrm{odd},\,\ell}\hat\sigma^x_{i,j,\ell}$ (for even $L_y$) reverses the sign
of $J_y$ and leaves $J_x$, $J_2$ and $\Jp$ unchanged. We therefore take $J_y>0$ and $\rp\ge0$ without
loss of generality, and write $r=|\rp|$.

\subsection{Classical limit and ordered structures}
\label{sec:classical}

At $g=0$ the in-plane coupling $J_x$ aligns neighboring spins along $x$, so in any low-energy
configuration each column along $y$ repeats in $x$, and the two layers are stacked in parallel for
$\rp>0$. The energy per spin is then fixed by the sequence of domains along $y$. For a structure in
which all domains have at least two sites and the wall density along $y$ is $\rho$ (walls per site),
\begin{equation}
e_0=-2+\kappa-\frac{r}{2}+(2-4\kappa)\,\rho ,
\label{eq:e0}
\end{equation}
in units of $J_0$. We denote a periodic structure by the lengths of the domains in one period,
$\langle n_1n_2\dots\rangle$; thus $\langle3\rangle$ is $\up\up\up\dn\dn\dn$ and $\langle23\rangle$ is
$\up\up\dn\dn\dn\up\up\dn\dn\dn\cdots$, which has period 5 and a net moment of $1/5$ per spin. The ferromagnet
(FM, $\rho=0$) has $e_0=-2+\kappa-r/2$, the antiphase
$\langle2\rangle$ ($\rho=1/2$) has $e_0=-1-\kappa-r/2$, and the structure $\langle3\rangle$,
$\up\up\up\dn\dn\dn$ ($\rho=1/3$), has $e_0=-4/3-\kappa/3-r/2$. At $\kappa=1/2$ the wall energy
$2-4\kappa$ vanishes and every structure with domains of at least two sites is degenerate. Because
walls must be straight lines along $x$, the number of such structures grows as $\exp(cL_y)$, not
$\exp(cN)$: the degeneracy is subextensive.

At zero temperature, the Suzuki--Trotter mapping~\cite{suzuki1976relationship} relates the
two-dimensional quantum model to an anisotropic three-dimensional classical ANNNI model. The thermal
lifting of the multiphase degeneracy of the classical model is known from its low-temperature
expansion~\cite{fisher1980infinitely,fisher1981low}, but the mapped model lies outside the regime of that
expansion (Sec.~\ref{sec:pt46}). At the
temperature of our simulations the thermodynamics is two-dimensional. The ferromagnet then orders
in the two-dimensional Ising universality class, and the structure $\langle n\rangle$, of period
$2n$, carries a $\mathbb Z_{2n}$ clock-type order; for $2n\ge6$ this order can melt through a floating
phase~\cite{villain1981two,pokrovsky1979ground}.
In mean-field theory the paramagnet becomes unstable at the wavevector $\cos\qs=1/(4\kappa)$ for
$\kappa>1/4$~\cite{selke1988annni}. This is the wavevector of the instability at the ordering
transition, not of the low-temperature ordered state.

\subsection{Observables}
\label{sec:observables}

We measure the energy per spin $e$, the transverse magnetization
$m_x=N^{-1}\sum\langle\hat\sigma^x\rangle$, and equal-time $zz$ correlations. Modulated order is
characterized by the per-layer axial amplitudes
\begin{equation}
a_\ell(q)=\frac{1}{L_xL_y}\sum_{i,j}\sigma^z_{i,j,\ell}\,e^{iqj},\qquad q=\frac{2\pi n}{L_y},
\label{eq:aq}
\end{equation}
and the structure factor $S(q)=\frac12\sum_\ell\langle|a_\ell(q)|^2\rangle$. The dominant wavevector
$\qs\in[0,\pi]$ maximizes $S(q)$. Since $a_\ell(-q)=a_\ell(q)^*$, the order parameter of a modulated
state with $0<\qs<\pi$ is $O(\qs)=2S(\qs)$, and of the ferromagnet $O(0)=S(0)$. Both saturate at 1
for the ferromagnet and for $\langle2\rangle$, at $8/9$ for $\langle3\rangle$, and at
$4(3+\sqrt5)/25\approx0.838$ for $\langle23\rangle$. We normalize by these saturation values when comparing
structures, and we identify the structure of a chain by $\qs$ together with the correlations along $y$
at distances 1 and 2. First-order transitions between ordered structures are located by crossings of
branch free energies (Sec.~\ref{sec:branches}). The Binder cumulant~\cite{binder1981finite}
$U_4=1-\langle m^4\rangle/(3\langle m^2\rangle^2)$ and the second-moment correlation length $\xi_y$
divided by the linear size are used only for the transverse-field Ising check of
Sec.~\ref{sec:validation}.

\section{Low-field perturbation theory near the multiphase point}
\label{sec:pt}

\subsection{Local fields}

For $g\ll1$ and temperatures far below the excitation gaps, the free energy of an ordered structure
is its energy~\eqref{eq:e0} dressed by virtual single-spin flips. A flip of spin $s$ costs
$2h J_0$, where $h$ is the local field in units of $J_0$. For a straight-wall structure with
favorable stacking,
\begin{equation}
h=2+r+a_{+1}+a_{-1}-\kappa\,(a_{+2}+a_{-2}),\qquad a_{d}=s_js_{j+d},
\label{eq:h}
\end{equation}
where $j$ is the coordinate along $y$ and the constant $2+r$ collects the two in-plane neighbors
along $x$ and the partner spin in the other layer. At second order,
\begin{equation}
e_2=-\frac{g^2}{2N}\sum_{\rm spins}\frac1h .
\label{eq:e2}
\end{equation}
At $\kappa=1/2$, Eq.~\eqref{eq:h} takes five values, listed in Table~\ref{tab:h} and illustrated in
Fig.~\ref{fig:model}(b). Every spin with $d\ge3$, and every spin of a two-site domain, has $h=3+r$. A spin next to a wall has the
smaller field $2+r$ and therefore gains more energy from quantum fluctuations. The field of a spin
depends only on the length of its own domain, provided every domain has at least two sites.

\begin{table}
\caption{Local field $h$ of Eq.~\eqref{eq:h} at $\kappa=1/2$ for a spin in a domain of $n$ sites, by its
distance $d$ from the nearer wall of the domain: $d=1$ for a spin next to a wall, $d=2$ for the next one, and so on.}
\label{tab:h}
\begin{ruledtabular}
\begin{tabular}{lccc}
Position & $d$ & $n$ & $h$ \\
\hline
Next to a wall & $1$ & $\ge3$ & $2+r$ \\
Second site from a wall & $2$ & $\ge4$ & $4+r$ \\
Bulk & $\ge3$ & $\ge5$ & $3+r$ \\
Two-site domain & $1$ & $2$ & $3+r$ \\
Center of a three-site domain & $2$ & $3$ & $5+r$ \\
\end{tabular}
\end{ruledtabular}
\end{table}

\subsection{The $\langle3\rangle$ wedge}
\label{sec:wedge}

\begin{figure*}
\includegraphics[width=\textwidth]{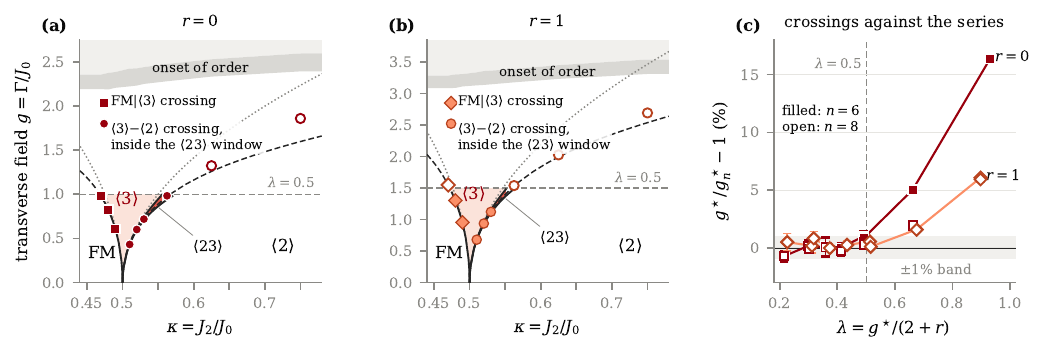}
\caption{Low-field phase diagram near the multiphase point for (a) $r=0$ and (b) $r=1$. Squares ($r=0$) and
diamonds ($r=1$): fields $g^\star$ at which the $\langle3\rangle$ branch crosses the FM branch ($\kappa<1/2$), a phase
boundary. Circles: crossings of the $\langle3\rangle$ and $\langle2\rangle$ branches ($\kappa>1/2$), which lie inside
the $\langle23\rangle$ window and are not phase boundaries. Both are measured on $12\times12$ bilayers
(Table~\ref{tab:crossings}); open symbols have $\lambda=g^\star/(2+r)>0.5$.
Solid lines: phase boundaries at $O(g^6)$ at the actual $\kappa$, drawn up to $\lambda=0.5$ (horizontal dashed
line). The shaded region is $\langle3\rangle$ and the red band $\langle23\rangle$ (Sec.~\ref{sec:pt46}); the
$\langle223\rangle$ window is too narrow to show. Dashed curves: $O(g^6)$ crossings of the $\langle3\rangle$ branch
with the FM and $\langle2\rangle$ branches, which the symbols measure. Dotted curves: the $O(g^2)$ wedge,
Eq.~\eqref{eq:wedge}. Grey band: onset of order along the anneals from the paramagnet ($L=12$), between the
fields at which the mean $S(\pi/3)$ rises through $0.1$ and $0.05$. (c) Deviation of the measured crossings
from the series at $O(g^6)$ (filled symbols) and $O(g^8)$ (open symbols) against $\lambda$; the grey band is
$\pm1\%$. The $O(g^8)$ series has no crossing at $\kappa=0.75$, $r=0$. Error bars are one standard error; in (a) and (b)
they are smaller than the symbols.}
\label{fig:wedge}
\end{figure*}

Measured from the ferromagnet, the energy per spin of a sequence of domains of lengths $n_j$ is
\begin{equation}
\Delta e=\frac{\sum_j\big[(2-4\kappa)-\tfrac{g^2}{2}F(n_j)\big]}{\sum_jn_j},
\label{eq:de}
\end{equation}
where $F(n)=\sum_{\rm domain}h^{-1}-n/(3+r)$ is the excess second-order gain of one domain. From
Table~\ref{tab:h},
\begin{align}
F(2)&=0,\qquad F(3)=\frac{6}{(2+r)(3+r)(5+r)},\nonumber\\
F(n\ge4)&=F(3)-\frac{2}{(3+r)(4+r)(5+r)} .
\label{eq:F}
\end{align}
Because Eq.~\eqref{eq:de} is a ratio of sums, its minimum over all sequences is reached by a periodic
structure $\langle n\rangle$ with a single domain length, the one that minimizes the per-domain
cost divided by $n$, or by the ferromagnet if every domain costs a positive energy
(Appendix~\ref{app:pt}). Since $F(3)>F(n\ge4)$ for every $r\ge0$, no structure with domains longer
than three is ever stable at this order. Comparing FM, $\langle3\rangle$ and $\langle2\rangle$ gives
the phase boundaries
\begin{equation}
\kappa_{\rm FM|3}=\frac12-\frac{g^2}{8}F(3),\qquad
\kappa_{3|2}=\frac12+\frac{g^2}{4}F(3).
\label{eq:wedge}
\end{equation}
Quantum fluctuations therefore open a wedge of $\langle3\rangle$ of width
\begin{equation}
\Delta\kappa=\frac{3g^2}{8}F(3)=\frac{9g^2}{4(2+r)(3+r)(5+r)}
\label{eq:width}
\end{equation}
that straddles $\kappa=1/2$, with twice as much of it on the antiphase side. For decoupled layers,
$F(3)=1/5$ and the wedge is $1/2-g^2/40<\kappa<1/2+g^2/20$. For $r=1$, $F(3)=1/12$ and the wedge is
$1/2-g^2/96<\kappa<1/2+g^2/48$, narrower by the factor $72/30=2.4$ at fixed $g$ [Fig.~\ref{fig:wedge}(a),(b)]. The
interlayer coupling adds the same constant $r$ to every local field, which compresses the differences between $1/h$
values that drive the selection. Most of the factor reflects the larger local fields: in terms of
$\lambda\equiv g/(2+r)$, the field in units of the local exchange scale, the wedge is $0.300\lambda^2$ wide for $r=0$
and $0.281\lambda^2$ for $r=1$, a ratio of $1.07$. The selection of $\langle3\rangle$ from a degenerate manifold is an instance of
order by disorder~\cite{villain1980order,shender1982antiferromagnetic,henley1989ordering}, here driven by
quantum fluctuations, as in other frustrated transverse-field Ising
models~\cite{moessner2000two,moessner2001ising}. These are the leading terms. At order $g^4$ (Sec.~\ref{sec:pt46}) the
$\langle3\rangle$ wedge widens by $0.0161\,g^4$ ($r=0$) and $0.0024\,g^4$ ($r=1$), and above
$\kappa=1/2$ it ends at a new structure, $\langle23\rangle$, rather than at $\langle2\rangle$.

The two boundaries of Eq.~\eqref{eq:wedge} behave differently at higher order. At
$\kappa_{\rm FM|3}$ every domain other than a three-site one costs energy at order $g^2$: a two-site
domain costs $\tfrac12g^2F(3)$ and a domain of $n\ge4$ sites $\tfrac12g^2[F(3)-F(n)]$. Higher orders
cannot overcome these costs at small $g$, so the transition between the ferromagnet and
$\langle3\rangle$ is a simple first-order level crossing. At $\kappa_{3|2}$, in contrast, every
sequence of two- and three-site domains has the same energy at order $g^2$, as in the thermal
problem~\cite{fisher1980infinitely}, and higher orders decide which of them are stable.
Section~\ref{sec:pt46} shows that the fourth order stabilizes $\langle23\rangle$ there, and the sixth
order $\langle223\rangle$.

\subsection{Fourth, sixth and eighth order}
\label{sec:pt46}

At $\kappa_{3|2}$ every sequence of two- and three-site domains has the same energy at order $g^2$, so
higher orders decide which structures are stable there. We carry the expansion to order $g^8$. Moving a
wall takes a whole row of flips, so structures with different walls are coupled only at order $g^{L_x}$
($g^{2L_x}$ for $r>0$). The effective Hamiltonian in the manifold of straight-wall structures is therefore
diagonal through order $g^{L_x-1}$ ($g^{2L_x-1}$ for $r>0$), and its diagonal elements are the nondegenerate Rayleigh--Schr\"odinger
energies of the individual structures. Taking $L_x\to\infty$ at fixed order, each straight-wall structure is a
nondegenerate reference state. We write its energy per spin as
\begin{equation}
e=e_0+\varepsilon_2g^2+\varepsilon_4g^4+\varepsilon_6g^6+\varepsilon_8g^8+O(g^{10}),
\label{eq:series}
\end{equation}
where $\varepsilon_2g^2$ is the term $e_2$ of Eq.~\eqref{eq:e2}. Odd orders vanish because every flipped
spin must be flipped back. Write the Ising part of Eq.~\eqref{eq:H} as
$-\sum J_{ij}\hat\sigma^z_i\hat\sigma^z_j$, so that $J_{ij}=-\kappa$ for the pairs at distance 2 along $y$.
Let $D_i=2h_i$ be the cost of flipping spin $i$, and let $B_{ij}=J_{ij}s_is_j$, which is minus the energy of
the bond between spins $i$ and $j$. Then
\begin{equation}
N\varepsilon_4=\sum_i\frac{1}{D_i^3}
-\sum_{\langle ij\rangle}\frac{4B_{ij}\,(D_i+D_j)}{(D_iD_j)^2\,(D_i+D_j-4B_{ij})} .
\label{eq:eps4}
\end{equation}
The first term is the $g^4$ term of an isolated spin, whose energy is $-(h^2+g^2)^{1/2}$. The second sum
runs over coupled pairs only. For an uncoupled pair, the fourth-order processes that flip both spins cancel
exactly against the normalization term of the expansion. This cancellation generalizes. Each spin
that takes part in a process is flipped and restored, so the $g^{2n}$ coefficient receives contributions
only from connected clusters of at most $n$ spins. We obtain $\varepsilon_6$ and $\varepsilon_8$ from this
linked-cluster form, solving every connected cluster of up to three and four spins exactly in rational
arithmetic.
Equation~\eqref{eq:eps4} is exact; exact diagonalization of small clusters reproduces it to
$3\times10^{-6}$ relative, the accuracy of the fit to the diagonalization data.
The sixth-order coefficient agrees with exact diagonalization to $10^{-4}$ relative,
and with an independent evaluation that does not use the cluster decomposition to $10^{-15}$.
The eighth-order coefficient reproduces the exact series of the transverse-field Ising chain,
$-25g^8/16384$ per spin, and equals a brute-force expansion over the full Hilbert space of 12-spin
clusters exactly. An independent calculation that does not use the cluster decomposition agrees with it to
twelve digits.
Table~\ref{tab:eps} lists the coefficients at $\kappa=1/2$.

\begin{table}
\caption{Coefficients of Eq.~\eqref{eq:series} per spin at $\kappa=1/2$, in units of $J_0$, for the
structures that compete near $\kappa_{3|2}$; $\rho$ is the wall density. $W_4=k\varepsilon_2'+\varepsilon_4$
with $k=F(3)/4$ [Eq.~\eqref{eq:W}] is the fourth-order energy on the $\langle3\rangle|\langle2\rangle$ line
(not needed for FM). $\varepsilon_2$ and $\varepsilon_4$ are exact; $W_4$, $\varepsilon_6$ and $\varepsilon_8$
are rounded (exact values are given with the data~\cite{truong2026zenodo}).}
\label{tab:eps}
\footnotesize
\begin{ruledtabular}
\begin{tabular}{llcccccc}
$r$ & & $\rho$ & $\varepsilon_2$ & $\varepsilon_4$ & $10^3\,W_4$ & $10^5\,\varepsilon_6$ & $10^5\,\varepsilon_8$ \\
\hline
0 & FM                    & 0   & $-1/6$ & $-5/1512$            & ---      & $-24.15$ & $-6.832$ \\
  & $\langle3\rangle$     & 1/3 & $-1/5$ & $-1/125$             & $-7.333$ & $-7.454$ & $+60.27$ \\
  & $\langle23\rangle$    & 2/5 & $-14/75$ & $-54673/7560000$     & $-4.610$ & $-120.5$ & $-30.38$ \\
  & $\langle2\rangle$     & 1/2 & $-1/6$ & $-13/4320$           & $2.546$  & $-22.06$ & $-5.044$ \\
1 & FM                    & 0   & $-1/8$ & $-1/1152$            & ---      & $-2.062$ & $-0.2020$ \\
  & $\langle3\rangle$     & 1/3 & $-5/36$ & $-437/362880$        & $-1.011$ & $4.409$  & $-0.1433$ \\
  & $\langle23\rangle$    & 2/5 & $-2/15$ & $-4477/3628800$      & $-0.5972$ & $-2.175$ & $-0.5922$ \\
  & $\langle2\rangle$     & 1/2 & $-1/8$ & $-11/13440$          & $0.4836$ & $-2.055$ & $-0.1779$ \\
\end{tabular}
\end{ruledtabular}
\end{table}

\paragraph*{Expansion about the multiphase point.}
Write $\kappa=\tfrac12+kg^2+k_1g^4+k_2g^6+k_3g^8$. The classical energy~\eqref{eq:e0} is linear in $\kappa$ with
slope $1-4\rho$, so Eq.~\eqref{eq:series} becomes
\begin{align}
e={}&e_0\big|_{\kappa=1/2}+g^2\big[k(1-4\rho)+\varepsilon_2\big]\nonumber\\
&+g^4\big[k_1(1-4\rho)+W_4\big]+g^6\big[k_2(1-4\rho)+W_6\big]\nonumber\\
&+g^8\big[k_3(1-4\rho)+W_8\big]+O(g^{10}),
\label{eq:expand}
\end{align}
with the $\kappa$ derivatives (primes) taken at $\kappa=1/2$:
\begin{align}
W_4={}&k\varepsilon_2'+\varepsilon_4,\qquad
W_6=k_1\varepsilon_2'+\tfrac12k^2\varepsilon_2''+k\varepsilon_4'+\varepsilon_6,\nonumber\\
W_8={}&k_2\varepsilon_2'+kk_1\varepsilon_2''+\tfrac16k^3\varepsilon_2'''+k_1\varepsilon_4'+\tfrac12k^2\varepsilon_4''
+k\varepsilon_6'+\varepsilon_8 .
\label{eq:W}
\end{align}
where $W_6$ and $W_8$ depend on the boundary considered through $k_1$ and $k_2$ (for instance $k_1^-$ or $k_1^+$ of
Eq.~\eqref{eq:w23}). On the second-order boundary, $k=F(3)/4$, the $g^2$ bracket is the same for every sequence of two- and
three-site domains (Appendix~\ref{app:pt}). Each domain of $n\ge4$ sites adds $\tfrac12g^2[(n-2)F(3)-F(n)]>0$
to the numerator of Eq.~\eqref{eq:de}.
At order $g^4$ only the sequences of 2s and 3s therefore compete, with energy
$k_1(1-4\rho)+W_4$ per spin. Since the first term is linear in $\rho$, the stable structures are the vertices
of the lower convex hull of $W_4$ plotted against $\rho$. The same construction with $W_6$ and $W_8$ decides
the boundaries that remain degenerate at the preceding order. The $\kappa$ derivatives are obtained exactly by
carrying $\kappa=\tfrac12+\delta$ through the cluster sums as a truncated power series in $\delta$.

\paragraph*{Fourth order: the $\langle23\rangle$ phase.}
At $\kappa=1/2$ the field~\eqref{eq:h} of a spin depends only on its position in its own domain
(Table~\ref{tab:h}). The pairs in Eq.~\eqref{eq:eps4} lie either within one domain or on the two sides of a
single wall. $W_4$ is therefore a sum of terms for single domains and for pairs of adjacent domains
[Eq.~\eqref{eq:ab}]. A periodic sequence can be read as a closed path through the domain lengths, and its
energy per spin is the cost of the path divided by its length. Such a ratio is smallest on an elementary
cycle, so only $\langle2\rangle$, $\langle3\rangle$ and $\langle23\rangle$ ($\up\up\dn\dn\dn$) can be stable at this
order, among sequences of any length (Appendix~\ref{app:pt}). Table~\ref{tab:eps} places $\langle23\rangle$
strictly below the chord from $\langle3\rangle$ to $\langle2\rangle$, so $\langle23\rangle$ is stable in a window
\begin{equation}
\tfrac12+\tfrac14F(3)g^2+k_1^-g^4<\kappa<\tfrac12+\tfrac14F(3)g^2+k_1^+g^4,
\label{eq:w23}
\end{equation}
with the coefficients of Table~\ref{tab:bounds}.
Its width, $\Delta\kappa_{23}=(k_1^+-k_1^-)g^4$, is 6.7 times smaller in the bilayer than in the single layer at
fixed $g$, against 2.4 for the $\langle3\rangle$ wedge. Most of this difference is power counting, $(3/2)^4$ against
$(3/2)^2$: at fixed $\lambda$ the two ratios are $1.32$ and $1.07$. At fixed $\lambda$ the window still narrows steadily
with $r$, from $0.123\lambda^4$ at $r=0$ to $0.107\lambda^4$, $0.093\lambda^4$ and $0.071\lambda^4$ at $r=1/2$, $1$ and
$2$. All these factors are ratios of asymptotic coefficients.
The metastable branches $\langle3\rangle$ and $\langle2\rangle$ cross inside the window, at
$\kappa=\tfrac12+\tfrac14F(3)g^2+k_1^{32}g^4$, the line along which their template free energies can be compared with
simulations. On the ferromagnetic side the boundary moves to $\kappa_{\rm FM|3}=\tfrac12-\tfrac18F(3)g^2+k_1^{\rm F}g^4$,
and no new structure appears there: every other structure contains two-site domains or domains of four or more sites,
each of which costs energy at order $g^2$ that higher orders cannot overturn at small $g$.

\begin{table}
\caption{Coefficients of the phase boundaries near $\kappa=1/2$. The edges of the $\langle23\rangle$ window are
$\kappa=\tfrac12+\tfrac14F(3)g^2+k_1^\pm g^4+k_2^\pm g^6$ (from sixth order the edge on the $\langle2\rangle$
side is the $\langle23\rangle|\langle223\rangle$ boundary); $k_1^{32}$ and $k_1^{\rm F}$ belong to the crossing of the
$\langle3\rangle$ and $\langle2\rangle$ branches and to the FM$|\langle3\rangle$ boundary. $\Delta\kappa_s$ is the width
of the window of structure $s$ at the order at which it opens, and $\delta_{233}^{(6)}g^6+\delta_{233}^{(8)}g^8$ the
height of $\langle233\rangle$ per spin above the segment from $\langle3\rangle$ to $\langle23\rangle$. Decimal entries are rounded.}
\label{tab:bounds}
\begin{ruledtabular}
\begin{tabular}{lll}
 & $r=0$ & $r=1$ \\
\hline
$k_1^-$ & $20591/2016000$ & $167/107520$ \\
$k_1^+$ & $6011/336000$ & $1961/725760$ \\
$k_1^{32}$ & $1067/72000$ & $31/13824$ \\
$k_1^{\rm F}$ & $-59/10080$ & $-1573/1935360$ \\
$10^3\,k_2^-$ & $-3.06$ & $-0.143$ \\
$10^3\,k_2^+$ & $+5.26$ & $+0.242$ \\
$\Delta\kappa_{23}/g^4$ & $7.68\times10^{-3}$ & $1.15\times10^{-3}$ \\
$\Delta\kappa_{223}/g^6$ & $4.1\times10^{-4}$ & $2.3\times10^{-5}$ \\
$\Delta\kappa_{2223}/g^8$ & $1.8\times10^{-5}$ & $3.8\times10^{-7}$ \\
$\delta_{233}^{(6)}$ & $+2.0\times10^{-6}$ & $+1.2\times10^{-7}$ \\
$\delta_{233}^{(8)}$ & $-9.7\times10^{-7}$ & $-1.8\times10^{-8}$ \\
\end{tabular}
\end{ruledtabular}
\end{table}

\paragraph*{Degenerate edges.}
The argument also shows exactly which structures tie on the window edges. In the $(\rho,W_4)$ plane, a
sequence with no two adjacent three-site domains ($\langle223\rangle$, $\langle2223\rangle$,
$\langle22323\rangle,\dots$) lies exactly on the segment joining $\langle2\rangle$ and $\langle23\rangle$. A
sequence with no two adjacent two-site domains ($\langle233\rangle$, $\langle2333\rangle,\dots$) lies exactly on
the segment joining $\langle23\rangle$ and $\langle3\rangle$ (Appendix~\ref{app:pt}).
Each edge of the window is therefore itself a multiphase line at order $g^4$, and the next order decides it.

\paragraph*{Sixth order.}
A connected cluster of three spins spans at most one intermediate domain. $W_6$ is therefore a sum of terms
for three consecutive domains, which we have confirmed in exact arithmetic.
Applied to pairs of consecutive domains, the elementary-cycle argument allows at most one new structure in
each gap at this order: $\langle223\rangle$ between $\langle23\rangle$ and $\langle2\rangle$, or
$\langle233\rangle$ between $\langle3\rangle$ and $\langle23\rangle$. $\langle223\rangle$ opens
(Table~\ref{tab:bounds}).
$\langle233\rangle$ does not, but it lies above the segment from $\langle3\rangle$ to $\langle23\rangle$ by only
about $2\times10^{-6}g^6$ per spin at $r=0$ (Table~\ref{tab:bounds}), so the edge is only weakly first order; beyond the
range of the series the simulations find $\langle233\rangle$ and $\langle2333\rangle$ below both
(Sec.~\ref{sec:res_beyond}). Every other
mixture of the two ($\langle2333\rangle$, $\langle23233\rangle$ and longer ones) contains at least one triple $(2,3,3)$
or $(3,3,2)$ of consecutive domains and also lies above the segment, so the $\langle3\rangle|\langle23\rangle$ boundary is
of first order at small $g$.

\paragraph*{Eighth order.}
At order $g^8$ the energy is a sum of terms for four consecutive domains. On the $\langle223\rangle|\langle2\rangle$
line $\langle2223\rangle$ lies below the chord from $\langle223\rangle$ to $\langle2\rangle$ and opens a narrow window
(Table~\ref{tab:bounds}). Longer mixtures such as $\langle22223\rangle$ lie on the new segments, so $\langle2223\rangle$ is
the only structure that enters this line at this order. On the $\langle23\rangle|\langle223\rangle$ line
the elementary cycles are $\langle23\rangle$ and $\langle223\rangle$ alone (Appendix~\ref{app:pt}), so no structure enters
there at this order; mixtures such as $\langle23223\rangle$ remain exactly degenerate, and a higher order decides that
edge. On the
$\langle3\rangle|\langle23\rangle$ line the eighth order lowers the margin of $\langle233\rangle$, but the truncated series
would change its sign only at $g\approx1.42$ ($r=0$) and $2.60$ ($r=1$), beyond the range in which it converges (see
below).
Through order $g^8$ the transverse field thus selects $\langle3\rangle$, $\langle23\rangle$, $\langle223\rangle$ and
$\langle2223\rangle$, in this order, between the ferromagnet and the antiphase. A structure $\langle2^k3\rangle$ can
first appear at order $g^{2(k+1)}$, the lowest order at which a connected cluster spans the $k+1$ domains of its
period, and it does so for $k\le3$. Whether it does so for every $k$ depends on the sign of each new coefficient;
in the classical expansion, corrections at large $k$ make the sequence finite at any nonzero temperature and admit
mixed phases~\cite{yeomans1988theory,selke1988annni}. Higher orders would decide this here (Sec.~\ref{sec:conclusion}).

The same order of appearance occurs in the low-temperature expansion of the classical three-dimensional
model~\cite{fisher1980infinitely,fisher1981low,selke1988annni}. There $\langle3\rangle$ appears at first order in
the expansion parameter with a first-order boundary to the ferromagnet, $\langle23\rangle$ at second order, and
each $\langle2^k3\rangle$ at order $k+1$, the number of connected spins that must flip to span the
sequence~\cite{yeomans1988theory}. In the language of domain-wall
interactions~\cite{szpilka1986domain,fisher1987domainI,fisher1987domainII}, $g^2$ takes over the role of the
low-temperature expansion parameter. The selecting excitations are virtual rather than thermal spin flips next to
the walls, and the coefficients are different. One difference is structural. In the classical calculation to second order, the
$\langle3\rangle|\langle23\rangle$ transition is of first order, while on the $\langle23\rangle|\langle2\rangle$ boundary all
sequences of two-bands and three-bands surrounded by two-bands remain
degenerate~\cite{fisher1981low}\cite[Sec.~4.2]{selke1988annni}; here both edges of the $\langle23\rangle$ window are
degenerate at $O(g^4)$, and the $\langle3\rangle|\langle23\rangle$ edge becomes first order only at $O(g^6)$. At $T=0$ the
quantum model maps onto a strongly anisotropic three-dimensional classical model~\cite{suzuki1976relationship}, outside
the regime of the classical low-temperature expansion. The classical boundary between the ferromagnet and
$\langle3\rangle$ has been derived rigorously at low temperature~\cite{dinaburg1985analysis}, and quantum Pirogov--Sinai
theory controls quantum perturbations of classical ground states, including infinitely degenerate
ones~\cite{datta1996lowI,datta1996lowII}. Our statements are asymptotic in the same sense: each holds at its order for
sufficiently small $g$, with $L_x\to\infty$ taken first (Sec.~\ref{sec:chain}).

The sixth order also shifts the edges of the $\langle23\rangle$ window (Table~\ref{tab:bounds}) and widens it by a
relative amount $1.08\,g^2$ at $r=0$ and $0.34\,g^2$ at $r=1$, as large as the leading term near $g=1$ for $r=0$. The
fourth-order width is therefore quantitative only at small $g$. At $r=0$ the eighth order is again large: $\varepsilon_8$
of $\langle3\rangle$ is eight times $\varepsilon_6$ in magnitude, with opposite sign (Table~\ref{tab:eps}).

\paragraph*{Crossings at finite $\kappa-1/2$.}
For fixed $\kappa>1/2$ the leading-order result, Eq.~\eqref{eq:wedge}, places the crossing of the
$\langle3\rangle$ and $\langle2\rangle$ branches at $g_{3|2}\simeq[4(\kappa-1/2)/F(3)]^{1/2}$ for $\kappa$
close to $1/2$.
Away from $\kappa=1/2$ the crossing is better obtained from Eq.~\eqref{eq:series} evaluated at the actual $\kappa$.
Table~\ref{tab:crossings} lists these crossings at each order next to the simulations (Sec.~\ref{sec:res_branches}).

\paragraph*{Convergence.}
The spins next to a wall of a domain of three or more sites feel no field from the couplings along $y$, at
any $\kappa$ [Eq.~\eqref{eq:h}]: their field $2+r$ comes from the neighbors along $x$ and from the partner in
the other layer. With its environment frozen, the row next to a wall is therefore a transverse-field Ising
chain for $r=0$ and a two-leg ladder with equal leg and rung couplings for $r=1$. Exact diagonalization of
chains of up to 16 sites and ladders of up to 10 rungs places their critical fields at $g=1.00$ and $1.83$, that is, at
$\lambda=0.50$ and $0.61$. The estimate ignores the weak pinning of the rows by the wall
interactions, so it identifies a scale rather than a radius of convergence, but the series for structures with
such rows is expected to deteriorate near these fields, where the rows next to the walls, and with them the walls,
soften. This is the quantum counterpart of the wall delocalization that destabilizes $\langle3\rangle$ in the
classical model~\cite{selke1988annni}. A breakdown of this kind need not show up in the low-order coefficients:
for $r=1$ at $\kappa=0.625$ and $0.75$ the $O(g^6)$ and $O(g^8)$ crossings nearly coincide, yet the simulated
crossings, which lie beyond the ladder scale, are well above both (Table~\ref{tab:crossings}). The edge rows set the range $\lambda\lesssim0.5$ ($r=0$) and $\lambda\lesssim0.6$ ($r=1$)
in which Sec.~\ref{sec:results} compares the series with the simulations.

\subsection{Why the square lattice differs from the chain}
\label{sec:chain}

The perturbative argument relies on the walls being rigid. On the square lattice, moving a wall by
one site requires flipping a whole row of $L_x$ spins (both rows, $2L_x$ spins, in the bilayer with
$r>0$), which enters only at order $g^{L_x}$ ($g^{2L_x}$), so the walls are frozen at any finite
order. In the chain ($J_x=0$, $r=0$) the field next to a wall of a domain of three or more sites vanishes by
Eq.~\eqref{eq:h}; at $\kappa=1/2$, where walls cost nothing, a single flip then moves a wall at no cost, and the walls delocalize, as described by
effective domain-wall Hamiltonians for the transverse chain~\cite{rieger1996one}. The degeneracy is
then lifted already at first order in $g$, and a gapless floating region opens between two lines
linear in $g$, $-g/2<\kappa-1/2<g$~\cite{chandra2007floating}, in place of the commensurate structures
found here. The large-spin analysis of an anisotropic Heisenberg model on the cubic
lattice~\cite{harris1995quantum,harris1995quantumb} also finds that quantum fluctuations lift the
degeneracy. There, however, the selecting excitations are spin deviations that propagate through the
transverse exchange. The resulting phases, $\langle3\rangle,\langle4\rangle,\langle5\rangle,\dots$, lie between
the antiphase and the ferromagnet with the period increasing toward the ferromagnet, and the
$\langle2\rangle|\langle3\rangle$ boundary remains of first order, with no $\langle23\rangle$ phase.
At the simulation temperature $T=0.15\,J_0$, thermal steps on a wall cost at least $2J_0$ each, and on a
periodic wall they come in pairs; even the conservative factor $e^{-2J_0/T}\approx10^{-6}$ keeps the walls
straight on the scale of our lattices. Virtual flips of the rows next to the walls are the relevant quantum
fluctuations, and they limit the convergence of the series (Sec.~\ref{sec:pt46}). As these rows approach their critical
field their kink energy falls, and thermal kinks, which do not move the walls, add to the free energy; Sec.~S3 of the Supplemental
Material~\cite{supp} estimates this contribution. The data agree. At $g_0=0.05$ the nearest-neighbor correlation along $x$ lies within
$6\times10^{-4}$ of unity on every template branch, the reduction expected from virtual single flips
alone ($4g_0^2\langle D^{-2}\rangle$ at second order). Steps on the walls would lower it further, and
they are not resolved.

The choice $J_x=J_y$ is not essential. For $J_x\neq J_y$, with energies in units of $J_y$, the constant $2+r$ in
Eq.~\eqref{eq:h} becomes $2J_x/J_y+r$, so every local field shifts by $2(J_x/J_y-1)$, and the second-order results,
Eqs.~\eqref{eq:wedge} and \eqref{eq:width}, hold with $r$ replaced by $r+2(J_x/J_y-1)$. At higher orders $J_x$ also enters the couplings between flipped spins, so the coefficients
change, but the rigidity of the walls for any $J_x>0$, the dependence of each local field on its own domain only, and
the elementary-cycle argument do not; the same candidates arise at each order, and whether they open depends on the
coefficients. For $r=0$ the rows next to the walls become critical at $\Gamma\approx J_x$, so the range of the series
shrinks as $J_x\to0$, toward the chain limit described above, where the walls move freely.

\section{Methods}
\label{sec:methods}

\subsection{Stochastic series expansion}

We write Eq.~\eqref{eq:H} as
$\hat H=-\sum_b\hat H_b-\sum_s(\hat H_{s,c}+\hat H_{s,f})+C$, with bond operators
$\hat H_b=|J_b|+J_b\hat\sigma^z_i\hat\sigma^z_j$, where $J_b$ is the coupling of bond $b$ in the
convention $-J_b\hat\sigma^z\hat\sigma^z$, a constant site operator $\hat H_{s,c}=\Gamma$, a spin-flip
operator $\hat H_{s,f}=\Gamma\hat\sigma^x_s$, and $C=\sum_b|J_b|+N\Gamma$. All matrix elements are
nonnegative for either sign of $J_b$, so frustration affects only the efficiency. Diagonal updates
insert and remove $\hat H_b$ and $\hat H_{s,c}$ in the operator string, and cluster updates
build Swendsen--Wang-type clusters that pass through bond operators and end on site
operators~\cite{sandvik2003stochastic}. The energy is $e=(-\langle n\rangle/\beta+C)/N$ with $n$ the
number of operators, the transverse magnetization is
$m_x=\langle n_f\rangle/(\beta\Gamma N)$ with $n_f$ the number of flip operators, and the
nearest-neighbor, next-nearest-neighbor and interlayer correlations follow from the numbers of
bond operators of each type. Equal-time quantities are measured on the propagated state at
imaginary time zero. There is no Trotter error. A Monte Carlo sweep consists of one diagonal update of the
whole operator string and one cluster update, in which every cluster is flipped with probability $1/2$. The
string length $M$ grows during thermalization, to $n+n/2+64$ whenever $n>3M/4$, and is fixed during
measurements. Unless stated otherwise, every chain performs 2000 thermalization and 3000 measurement sweeps
at each field, with measurements averaged in bins of 100 sweeps. At the test field of Sec.~\ref{sec:res_23}
($\kappa=0.53$, $r=1$, $g_c=1.12$) the integrated autocorrelation times, estimated from bins of 10 sweeps, are about 6
sweeps for the energy and $12$--$17$ sweeps for $m_x$, the same for $\langle3\rangle$ and $\langle23\rangle$ on
$12\times12$, $20\times30$ and $20\times60$ lattices. A field thus provides about 200 independent values of $m_x$ per
chain, and the 2000 thermalization sweeps exceed these times a hundredfold. Near the crossover of the 12-spin cluster
of Sec.~\ref{sec:validation} the time for $m_x$ grows to about 50 sweeps.

\subsection{Equilibration in the modulated region}
\label{sec:branches}

\begin{figure*}
\includegraphics[width=\textwidth]{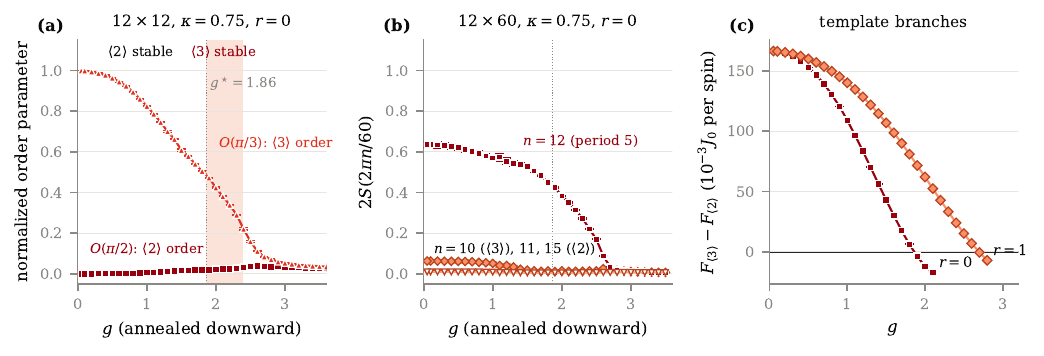}
\caption{Annealing freezes at the wavevector at which order forms. (a) Order parameters along anneals from the
paramagnet at $\kappa=0.75$, $r=0$ on $12\times12$ bilayers (means of six chains), normalized to their saturation values:
$O(\pi/3)$ of $\langle3\rangle$ (triangles) and $O(\pi/2)$ of $\langle2\rangle$ (squares). The shaded range is where
$\langle3\rangle$ lies below the FM and $\langle2\rangle$ branches, between their crossing $g^\star$
(Table~\ref{tab:crossings}) and the onset of order. Below $g^\star$ the antiphase is lower, yet the chains stay in
$\langle3\rangle$. (b) The same anneal on $12\times60$ bilayers, where $\langle3\rangle$ ($n=10$ in $q=2\pi n/60$), $n=11$,
period 5 ($n=12$) and $\langle2\rangle$ ($n=15$) are resolved: order forms at $n=12$, the allowed wavevector nearest the
mean-field instability, and stays there. The dotted line marks $g^\star$. (c) $F_{\langle3\rangle}-F_{\langle2\rangle}$ from
the template branches at $\kappa=0.75$, which changes sign at $g^\star$.}
\label{fig:anneal}
\end{figure*}

At small $g$ the operator string holds few site operators, the clusters of satisfied bonds percolate, and the
cluster update reduces to global flips, so local moves cannot convert one ordered structure into another. On 12-site
clusters, annealing downward from the paramagnet removes these traps. On larger lattices it does not.
Figure~\ref{fig:anneal}(a) shows an anneal on a $12\times12$ bilayer at $\kappa=0.75$: $\langle3\rangle$ order forms at
the ordering transition and persists below the field $g^\star$ at which $\langle2\rangle$ becomes lower, so at $r=0$ the chains end
$0.17J_0$ per spin above the stable antiphase. Of 138 chains annealed to $g_0=0.05$ in the runs of
Sec.~\ref{sec:res_branches}, at $\kappa$ between $0.47$ and $0.75$ and on $12\times12$ and $24\times24$ lattices, 137 end in
$\langle3\rangle$, even for $\kappa<1/2$ where the ferromagnet is stable; the remaining one ends in another sequence with
the same wall density. Annealing ten times more slowly leaves this unchanged on $12\times12$ lattices, where only $\langle2\rangle$ and
$\langle3\rangle$ fit: at $\kappa=0.75$ all twelve chains, six for each $r$, again end in $\langle3\rangle$.

This freezing is set by the wavevector at which order forms. On $L_y=12$ the allowed wavevector nearest the mean-field
instability, $\cos\qs=1/(4\kappa)$, is $\pi/3$ for every $\kappa$ studied, and only $\langle2\rangle$ and $\langle3\rangle$ fit
among sequences of two- and three-site domains. On $12\times60$ lattices, which resolve the wavevectors in between, order
forms instead at the allowed wavevector nearest the instability and stays there [Fig.~\ref{fig:anneal}(b)]; at
$\kappa=0.625$ and $0.75$ every chain ends in a sequence of two- and three-site domains with that wall density, none in
$\langle3\rangle$. In
separate low-field checks on $24\times24$ lattices, anneals from $g=3$ ($r=0$) or $5$ ($r=1$) also leave frozen stripes at
$\kappa=0.25$, as in zero-temperature quenches of Ising ferromagnets~\cite{spirin2001fate}, and frozen domain-wall
configurations for $\kappa\ge0.75$. Exchange Monte Carlo
in the field~\cite{hukushima1996exchange} could reduce the freezing; we do not pursue it, because the template branches
below give the free energies directly.

We therefore treat each candidate ordered structure as a separate branch. Chains are started in the FM,
$\langle2\rangle$ and $\langle3\rangle$ structures at $g_0$ and stepped upward in $g$. Along each branch the free energy
per spin follows from $\partial F/\partial\Gamma=-\langle\sum_s\hat\sigma^x_s\rangle$,
\begin{equation}
f_b(g)=f_b(g_0)-\int_{g_0}^{g}m_x^{(b)}(g')\,dg' .
\label{eq:ti}
\end{equation}
At $g_0$ the entropy of an ordered state is negligible (the lowest excitation costs $2(2+r)J_0$), so we take $f_b(g_0)$
from the perturbation series~\eqref{eq:series} of the template; the measured energies at $g_0$ agree with it
[Fig.~\ref{fig:validation}(b)], and using the series instead of the measured starting point reduces the uncertainty of
the crossings by a median factor of $3.6$; crossings from the measured starting points agree within $2\sigma$. The branch free energy is that of one fixed arrangement of walls. The entropy
of the many arrangements with similar energies, of order $1/L_x$ per spin, is not sampled because the walls are rigid. It
could reach $10^{-3}J_0$ per spin on our lattices, so the Gibbs state of a finite lattice would mix arrangements, but it
vanishes in the thermodynamic limit, where the branches are the appropriate objects to compare. Before the first cluster
update the operator string is filled by diagonal updates alone, so that the initial structure survives
(Appendix~\ref{app:sse}).

The integral is valid while a branch remains in one structure, so we cut a chain at the first change of its dominant
wavevector; every crossing lies at least $0.24$ below the first cut, and truncating the branches there moves the
crossings of Table~\ref{tab:crossings} by at most $3\times10^{-4}$. We integrate $m_x$ with the trapezoidal rule and a Richardson correction, locate crossings by a
cubic-spline root, and estimate errors by the jackknife over independent chains, with Student-$t$ intervals; these choices were fixed after a first
analysis of the data of Sec.~\ref{sec:res_branches} and before the test of Sec.~\ref{sec:res_23}. The stable state at each $g$ is the branch of
lowest $f_b$ among the structures compared, and a crossing of two branches locates a first-order transition; how far a branch persists beyond it depends on the lattice and the update and
is not a thermodynamic hysteresis. As a check of the integration, the measured energy $e_b$ must equal $f_b$ along the
branches; the differences scatter as expected from the chain statistics, although this check cannot detect an offset
that shifts both together.

\subsection{Validation}
\label{sec:validation}

We benchmark the SSE code against exact diagonalization at finite temperature on four 12-spin
periodic clusters built by the production lattice code, so that both methods use the same bond list: a $2\times3$
bilayer, $2\times6$ and $3\times4$ single layers, and a $1\times6$ two-leg ladder, at parameters that include frustrated
points up to $\kappa=1.3$, both signs of $\rp$, $J_y<0$, and $T=0.15$ and $0.5\,J_0$, with longer runs
than in production (Table~S1 of the Supplemental Material~\cite{supp}). The 471
comparisons of the energy, $m_x$, both correlation estimators, template moments and the structure factor scatter
slightly less than the Student-$t$ distribution expected for our 24-chain averages [Fig.~\ref{fig:validation}(a)],
since comparisons within one parameter set share their chains; the largest $|z|$ is $2.62$.

We also test the whole free-energy procedure against exact diagonalization. On two 12-spin clusters, a $2\times3$
bilayer at $\kappa=0$, $r=1$ and a $2\times6$ layer at $\kappa=0.3$, chains started in the ferromagnet at $g_0$
are stepped up to $g=3$ exactly as the template branches, and $m_x$ is integrated with the same quadrature from
the exact free energy at $g_0$. On the bilayer the integral reproduces the exact $F(g)$ at every field within
$3\times10^{-4}$. On the frustrated layer, four runs that separate seeds from sweeps agree with the exact values within
$4\times10^{-4}$ below the crossover of this small cluster ($g\le1.2$); across the crossover two of them, with the same
seeds, lie above it by up to $2\times10^{-3}$. The integration is thus confirmed at the level of $4\times10^{-4}$ in the
ordered regime, where all comparisons of this paper lie, while across a crossover occasional errors of order $10^{-3}$
occur that more sweeps do not reliably remove.

On large lattices the energies of all template branches at $g_0$ agree with the perturbation
series~\eqref{eq:series} [Fig.~\ref{fig:validation}(b)]. For the two-dimensional transverse-field Ising model
($\kappa=\rp=0$) with $\beta=L$ (dynamic exponent $z=1$), the crossings of the Binder cumulant $U_4$ and of $\xi_y/L$
bracket the quantum critical point $g_c=3.04438(2)$~\cite{blote2002cluster} from below and above, within $0.2\%$ for the
largest pair [Fig.~\ref{fig:validation}(c)]. At $T=0.15\,J_0$ the transition lies between $g=3.0$, where
$T_c=0.2977(9)\,J_0$~\cite{hesselmann2016thermal}, and $g_c$. Our crossings at this temperature drift downward into this
interval but have not converged by $L=48$, where $U_4$ at the crossing still rises [Fig.~\ref{fig:validation}(d)]: the
crossover from quantum-critical to two-dimensional classical scaling, at $L\approx v\beta$, sets in only at these sizes.
For this reason we do not locate the paramagnetic boundary of the full model at $T=0.15$ in this work.

\section{Results}
\label{sec:results}

\subsection{The $\langle3\rangle$ wedge from branch free energies}
\label{sec:res_branches}

We computed template branches for the FM, $\langle2\rangle$ and $\langle3\rangle$ structures at nine values of $\kappa$
between $0.47$ and $0.75$, for $r=0$ and $1$, on $12\times12$ bilayers at $T=0.15\,J_0$, with six chains per branch and
field steps of $0.1$, and located their crossings as described in Sec.~\ref{sec:branches}. Table~\ref{tab:crossings} and
Fig.~\ref{fig:wedge} compare the crossing fields $g^\star$ with the perturbation series evaluated at the actual $\kappa$.
We use $\lambda=g^\star/(2+r)$ (Sec.~\ref{sec:wedge}) to measure how far the series is pushed.

\begin{table*}
\caption{Crossing fields of the $\langle3\rangle$ branch with the FM branch ($\kappa<1/2$) or the
$\langle2\rangle$ branch ($\kappa>1/2$) on $12\times12$ bilayers at $T=0.15\,J_0$, with jackknife
standard errors, against the crossings of the perturbation series truncated at $O(g^n)$ at the same
$\kappa$. $\lambda=g^\star/(2+r)$; $z_n=(g^\star-g^\star_n)/\sigma$. For $\kappa>1/2$ these are crossings of
two branches that lie inside the $\langle23\rangle$ window, not phase boundaries. Daggers mark the
inverse-variance combination of the production run with a finer-grid rerun (production and rerun values:
$0.5952(48)$, $0.6062(13)$; $0.4378(37)$, $0.4290(25)$; $1.1258(49)$, $1.1200(36)$; $1.3253(15)$, $1.3215(22)$; $1.8597(33)$,
$1.8603(25)$). At $\kappa=0.75$, $r=0$
the $O(g^8)$ series has no crossing.}
\label{tab:crossings}
\begin{ruledtabular}
\begin{tabular}{llllllllrr}
$\kappa$ & $r$ & $g^\star$ (QMC) & $\lambda$ & $g^\star_2$ & $g^\star_4$ & $g^\star_6$ & $g^\star_8$ & $z_6$ & $z_8$ \\
\hline
$0.47$ & 0 & $0.9856(32)$ & 0.49 & 1.0397 & 0.9766 & 0.9770 & 0.9829 & $+2.6$ & $+0.8$ \\
$0.48$ & 0 & $0.8250(38)$ & 0.41 & 0.8631 & 0.8247 & 0.8251 & 0.8271 & $0.0$ & $-0.5$ \\
$0.49$ & 0 & $0.6054(13)^\dagger$ & 0.30 & 0.6210 & 0.6056 & 0.6058 & 0.6060 & $-0.2$ & $-0.4$ \\
$0.51$ & 0 & $0.4318(21)^\dagger$ & 0.22 & 0.4408 & 0.4348 & 0.4349 & 0.4349 & $-1.5$ & $-1.5$ \\
$0.52$ & 0 & $0.6008(32)$ & 0.30 & 0.6149 & 0.5995 & 0.5997 & 0.6000 & $+0.3$ & $+0.3$ \\
$0.53$ & 0 & $0.7190(58)$ & 0.36 & 0.7432 & 0.7175 & 0.7179 & 0.7187 & $+0.2$ & $+0.1$ \\
$0.5625$ & 0 & $0.9832(48)$ & 0.49 & 1.0312 & 0.9713 & 0.9733 & 0.9795 & $+2.1$ & $+0.8$ \\
$0.625$ & 0 & $1.3241(13)^\dagger$ & 0.66 & 1.3674 & 1.2525 & 1.2606 & 1.2990 & $+50$ & $+20$ \\
$0.75$ & 0 & $1.8601(20)^\dagger$ & 0.93 & 1.7550 & 1.5669 & 1.5986 & --- & $+132$ & --- \\
$0.47$ & 1 & $1.5484(26)$ & 0.52 & 1.5803 & 1.5327 & 1.5469 & 1.5468 & $+0.6$ & $+0.6$ \\
$0.48$ & 1 & $1.3017(41)$ & 0.43 & 1.3196 & 1.2918 & 1.2984 & 1.2985 & $+0.8$ & $+0.8$ \\
$0.49$ & 1 & $0.9542(57)$ & 0.32 & 0.9555 & 0.9449 & 0.9465 & 0.9465 & $+1.4$ & $+1.4$ \\
$0.51$ & 1 & $0.6794(51)$ & 0.23 & 0.6800 & 0.6757 & 0.6761 & 0.6761 & $+0.7$ & $+0.7$ \\
$0.52$ & 1 & $0.9367(57)$ & 0.31 & 0.9446 & 0.9336 & 0.9351 & 0.9351 & $+0.3$ & $+0.3$ \\
$0.53$ & 1 & $1.1221(29)^\dagger$ & 0.37 & 1.1373 & 1.1187 & 1.1222 & 1.1222 & $-0.0$ & $-0.1$ \\
$0.5625$ & 1 & $1.5390(42)$ & 0.51 & 1.5605 & 1.5159 & 1.5307 & 1.5308 & $+2.0$ & $+2.0$ \\
$0.625$ & 1 & $2.0258(23)$ & 0.68 & 2.0346 & 1.9473 & 1.9941 & 1.9946 & $+14$ & $+13$ \\
$0.75$ & 1 & $2.6923(48)$ & 0.90 & 2.5495 & 2.4079 & 2.5381 & 2.5416 & $+32$ & $+31$ \\
\end{tabular}
\end{ruledtabular}
\end{table*}

Where the series converges the agreement is quantitative. With the eighth order, the twelve crossings with
$\lambda\le0.5$ agree with the series with $\chi^2=7.2$; the leading order alone is off by a few percent. At the two
points with $\lambda\approx0.49$ for $r=0$ the truncation error is not controlled, since the last step of the series is
larger than the previous one; these crossings lie at the critical field of the rows next to the walls. Without them, $\chi^2=6.0$ for the ten
remaining crossings, and the slope means of Table~\ref{tab:slopes} change by less than $0.001$.
Beyond
$\lambda=0.5$ the crossings move above the series [Fig.~\ref{fig:wedge}(c)], and the eighth order does not close the
gap. We chose the range $\lambda\le0.5$ after inspecting Table~\ref{tab:crossings}; the edge rows of
Sec.~\ref{sec:pt46} give it a physical basis, and with their range for $r=1$, $\lambda\le0.61$, the fourteen crossings
included still agree ($\chi^2=11.6$).

With the wedge boundaries written as $|\kappa-1/2|=Ag^2+Bg^4$ and $B$ fixed at its $O(g^4)$ value, each crossing gives an
estimate of the leading coefficient $A$ (Table~\ref{tab:slopes}). The weighted means agree with the identically processed
series to about $1\%$ on both sides and for both $r$, so the $\langle3\rangle$ wedge of the lowest-order theory is
confirmed, including the factor $2.4$ by which the second layer narrows it at fixed $g$ ($2.43(2)$ and $2.41(2)$ on the two sides; this
ratio follows from the slopes and is not an independent test). Excluding the reruns described below changes the means
by less than $0.01$; extending the window beyond $\lambda=0.5$ makes the fits on the $\langle2\rangle$ side fail.

\begin{table}
\caption{Leading coefficients of the wedge boundaries, $A_i=|\kappa-1/2|/g^{\star2}-Bg^{\star2}$ with
$B$ fixed at its $O(g^4)$ value, relative to $A_{\rm th}=F(3)/8$ (FM side) or $F(3)/4$ ($\langle2\rangle$
side), for the crossings with $\lambda\le0.5$. Last two columns: the same extraction applied to the $O(g^6)$ and $O(g^8)$ crossings of
Table~\ref{tab:crossings}, which shows the bias of the extraction itself. Last rows:
weighted means (with the QMC weights).}
\label{tab:slopes}
\begin{ruledtabular}
\begin{tabular}{lllllll}
Side & $r$ & $\kappa$ & $\lambda$ & $A_i/A_{\rm th}$ & $O(g^6)$ & $O(g^8)$ \\
\hline
FM$|\langle3\rangle$ & 0 & 0.47 & 0.49 & $1.008(10)$ & 1.034 & 1.016 \\
FM$|\langle3\rangle$ & 0 & 0.48 & 0.41 & $1.016(12)$ & 1.016 & 1.009 \\
FM$|\langle3\rangle$ & 0 & 0.49 & 0.30 & $1.005(5)$ & 1.004 & 1.003 \\
$\langle3\rangle|\langle2\rangle$ & 0 & 0.51 & 0.22 & $1.018(11)$ & 1.002 & 1.001 \\
$\langle3\rangle|\langle2\rangle$ & 0 & 0.52 & 0.30 & $1.001(13)$ & 1.006 & 1.005 \\
$\langle3\rangle|\langle2\rangle$ & 0 & 0.53 & 0.36 & $1.007(21)$ & 1.012 & 1.009 \\
$\langle3\rangle|\langle2\rangle$ & 0 & 0.5625 & 0.49 & $1.007(15)$ & 1.039 & 1.019 \\
FM$|\langle3\rangle$ & 1 & 0.48 & 0.43 & $1.001(8)$ & 1.007 & 1.007 \\
FM$|\langle3\rangle$ & 1 & 0.49 & 0.32 & $0.983(13)$ & 1.002 & 1.002 \\
$\langle3\rangle|\langle2\rangle$ & 1 & 0.51 & 0.23 & $0.990(16)$ & 1.001 & 1.001 \\
$\langle3\rangle|\langle2\rangle$ & 1 & 0.52 & 0.31 & $1.000(15)$ & 1.004 & 1.004 \\
$\langle3\rangle|\langle2\rangle$ & 1 & 0.53 & 0.37 & $1.008(7)$ & 1.008 & 1.008 \\
\hline
FM$|\langle3\rangle$ & 0 & mean & & $1.007(4)$ \\
$\langle3\rangle|\langle2\rangle$ & 0 & mean & & $1.010(7)$ \\
FM$|\langle3\rangle$ & 1 & mean & & $0.996(7)$ \\
$\langle3\rangle|\langle2\rangle$ & 1 & mean & & $1.005(6)$ \\
\end{tabular}
\end{ruledtabular}
\end{table}

Two checks bear on the errors. Five cases were rerun with new seeds on finer grids; four were chosen in advance, and
$\kappa=0.49$, $r=0$ was added because its production value lay $2.2$ standard errors below the $O(g^6)$ crossing, so we
combine each rerun with its production value rather than replace it. The two differ with $\chi^2=9.4$ for five cases
after allowing for the Student-$t$ distribution of six-chain errors ($p=0.09$), a weak hint that these errors are
somewhat small; other ways of combining them move the slopes by less than $0.01$. Second, $24\times24$ bilayers at five
of the points ($\kappa=0.47$ and $0.53$ for both $r$, and $\kappa=0.625$ for $r=0$) reproduce the $12\times12$ crossings
within $0.24\%$.

\subsection{The $\langle23\rangle$ phase}
\label{sec:res_23}

We tested the fourth-order prediction with simulations whose design, analysis and pass criteria were fixed
before the runs (Appendix~\ref{app:prereg}). At seven points with $\lambda=g/(2+r)\le0.5$, four for $r=0$ and three
for $r=1$, we followed $\langle23\rangle$ and $\langle2\rangle$ on $20\times20$ bilayers and $\langle23\rangle$ and
$\langle3\rangle$ on $20\times30$ bilayers, so that each comparison is made on one lattice, with 12 chains per
structure. The test field $g_c$ is the $O(g^6)$ crossing of the $\langle3\rangle$ and $\langle2\rangle$ branches, where
the series predicts the largest margin for $\langle23\rangle$. A point passes if $\langle23\rangle$ lies below both
neighbors at $g_c$ by more than two standard errors (one-sided Welch $t$) and no chain has left its structure.

\begin{table*}
\caption{Pre-registered test of the $\langle23\rangle$ phase, at $\lambda=g_c/(2+r)$ between $0.30$ and $0.49$. $\Delta F_2$ and $\Delta F_3$
are the free-energy differences per spin between $\langle23\rangle$ and $\langle2\rangle$ (on $20\times20$) and
between $\langle23\rangle$ and $\langle3\rangle$ (on $20\times30$) at $g_c$, in units of $10^{-4}J_0$, with standard
errors and Welch $t$ values. $\Delta F_{\rm th}$ is the $O(g^6)$ prediction, the same for both at $g_c$.
$g_{2|23}$ and $g_{23|3}$ are the edges of the window where $\langle23\rangle$ is below $\langle2\rangle$ and
below $\langle3\rangle$, next to their $O(g^6)$ values. $w=\Delta\kappa/g_c^4$ (units of $10^{-3}$) is the width of the
window converted to $\kappa$ with the slope of the measured $\langle3\rangle|\langle2\rangle$ line.}
\label{tab:s5a}
\footnotesize\setlength{\tabcolsep}{3pt}
\begin{ruledtabular}
\begin{tabular}{llllllllll}
$\kappa$ & $r$ & $g_c$ & $\Delta F_2$ ($t$) & $\Delta F_3$ ($t$) & $\Delta F_{\rm th}$ &
$g_{2|23}$ (th) & $g_{23|3}$ (th) & $w$ (th) & Test \\
\hline
0.52   & 0 & 0.5997 & $-3.32(60)$ ($-5.6$)  & $-1.46(62)$ ($-2.4$)  & $-2.17$  & 0.5885(21) (0.5923) & 0.6084(36) (0.6123) & 11.3(2.4) (11.4) & pass \\
0.53   & 0 & 0.7179 & $-4.93(76)$ ($-6.5$)  & $-7.19(99)$ ($-7.3$)  & $-4.98$  & 0.7050(19) (0.7049) & 0.7553(51) (0.7418) & 17.8(1.9) (13.4) & pass \\
0.545  & 0 & 0.8525 & $-12.13(82)$ ($-14.9$) & $-14.96(100)$ ($-14.9$) & $-11.42$ & 0.8291(16) (0.8302) & 0.9256(52) (0.8990) & 22.4(1.3) (16.6) & pass \\
0.5625 & 0 & 0.9733 & $-23.04(86)$ ($-26.8$) & $-30.65(122)$ ($-25.2$) & $-22.19$ & 0.9384(13) (0.9396) & 1.1605(91) (1.0561)$^a$ & 37.6(1.6) (21.2) & pass \\
0.52   & 1 & 0.9351 & $-0.99(54)$ ($-1.8$)  & $-2.47(58)$ ($-4.2$)  & $-1.75$  & 0.9297(30) (0.9254) & 0.9577(52) (0.9512) & 1.72(37) (1.58) & fail$^b$ \\
0.53   & 1 & 1.1222 & $-4.31(66)$ ($-6.5$)  & $-4.90(34)$ ($-14.5$) & $-3.97$  & 1.1037(29) (1.1051) & 1.1604(27) (1.1525) & 2.17(15) (1.81) & pass \\
0.545  & 1 & 1.3369 & $-8.65(91)$ ($-9.5$)  & $-9.57(71)$ ($-13.4$) & $-8.98$  & 1.3082(30) (1.3070) & 1.4020(48) (1.3951) & 2.34(14) (2.20) & pass \\
\end{tabular}
\end{ruledtabular}
\raggedright\footnotesize{$^a$Above $g=1.064(15)$ the structure $\langle233\rangle$ lies below $\langle23\rangle$ (Sec.~\ref{sec:res_beyond}).
$^b$Passes in a follow-up with 24 new chains per structure, decided after this result (Sec.~\ref{sec:res_23}).}
\end{table*}

\begin{table}
\caption{The $O(g^8)$ predictions at the seven test points of Table~\ref{tab:s5a}, computed after the test: the
free-energy differences $\Delta F_2$ and $\Delta F_3$ at the pre-registered $g_c$, which is the $O(g^6)$ crossing of the
$\langle3\rangle$ and $\langle2\rangle$ branches (in $10^{-4}J_0$), and the window edges. At $\kappa=0.5625$, $r=0$ the eighth-order series has no $\langle23\rangle|\langle3\rangle$ crossing below
$g=1.6$.}
\label{tab:s5a8}
\begin{ruledtabular}
\begin{tabular}{llrrll}
$\kappa$ & $r$ & $\Delta F_2^{(8)}$ & $\Delta F_3^{(8)}$ & $g_{2|23}^{(8)}$ & $g_{23|3}^{(8)}$ \\
\hline
0.52 & 0 & $-2.21$ & $-2.32$ & 0.5922 & 0.6134 \\
0.53 & 0 & $-5.16$ & $-5.63$ & 0.7045 & 0.7460 \\
0.545 & 0 & $-12.12$ & $-14.01$ & 0.8291 & 0.9176 \\
0.5625 & 0 & $-24.18$ & $-29.74$ & 0.9373 & --- \\
0.52 & 1 & $-1.78$ & $-1.78$ & 0.9253 & 0.9515 \\
0.53 & 1 & $-4.08$ & $-4.09$ & 1.1046 & 1.1536 \\
0.545 & 1 & $-9.42$ & $-9.44$ & 1.3058 & 1.3994 \\
\end{tabular}
\end{ruledtabular}
\end{table}

\begin{figure*}
\includegraphics[width=\textwidth]{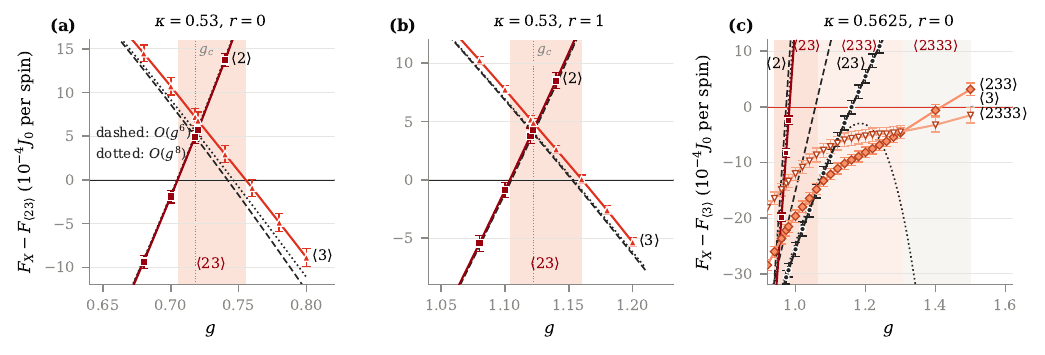}
\caption{Free energies of competing structures near the $\langle23\rangle$ window. (a), (b)
$F_X-F_{\langle23\rangle}$ at $\kappa=0.53$ for $r=0$ and $1$, for $X=\langle2\rangle$ (squares, $20\times20$
bilayers) and $X=\langle3\rangle$ (triangles, $20\times30$), each against $\langle23\rangle$ on the same lattice.
$\langle23\rangle$ is stable where both are positive (shaded, edges from the simulations). Dashed lines: $O(g^6)$,
dotted lines: $O(g^8)$, both at the actual $\kappa$. $g_c$ is the field of the pre-registered test. (c) $\kappa=0.5625$, $r=0$: free energies
relative to $\langle3\rangle$ ($20\times30$), including $\langle233\rangle$ ($20\times32$, diamonds) and
$\langle2333\rangle$ ($20\times22$, open triangles); $\langle23\rangle$ is shown as circles. The lowest curve is the
stable structure among those followed, and the shaded ranges mark where $\langle23\rangle$, $\langle233\rangle$ and
$\langle2333\rangle$ are lowest. In (b) the differences to $\langle3\rangle$ ($20\times30$, standard
sweeps) carry a bias of about $-10^{-4}$ (Sec.~\ref{sec:res_23}). Error bars are one standard error.}
\label{fig:f23}
\end{figure*}

\begin{figure*}
\includegraphics[width=\textwidth]{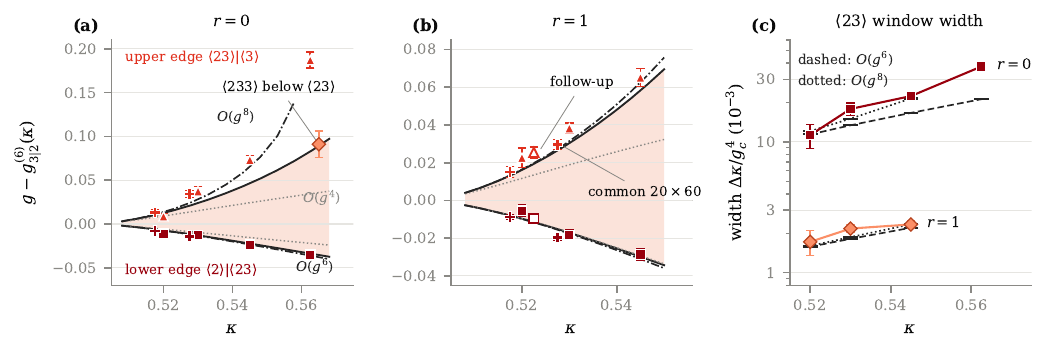}
\caption{Edges of the $\langle23\rangle$ window at the seven test points for (a) $r=0$ and (b) $r=1$, measured from
the $O(g^6)$ crossing $g^{(6)}_{3|2}(\kappa)$ of the $\langle3\rangle$ and $\langle2\rangle$ branches. Squares: lower
edge, where $\langle23\rangle$ drops below $\langle2\rangle$; triangles: upper edge, where $\langle3\rangle$ drops
below $\langle23\rangle$. Solid lines and shading: $O(g^6)$; dotted lines: $O(g^4)$; dash-dotted lines: $O(g^8)$,
whose upper edge for $r=0$ ends where the eighth-order series stops crossing. Open symbols at
$\kappa=0.52$, $r=1$: the follow-up pooled with the original chains (36 per structure), decided after the
pre-registered result. Plus
signs: edges with $\langle2\rangle$, $\langle23\rangle$ and $\langle3\rangle$ on one $20\times60$ lattice; for $r=1$ the
chains of two independent runs are pooled. Both are
displaced slightly in $\kappa$ for clarity. The
diamond at $\kappa=0.5625$, $r=0$ marks where $\langle233\rangle$ drops below $\langle23\rangle$
(Sec.~\ref{sec:res_beyond}). (c) Widths $w=\Delta\kappa/g_c^4$ of the windows (Table~\ref{tab:s5a}) against
$O(g^6)$ (dashed) and $O(g^8)$ (dotted; no value at $\kappa=0.5625$, $r=0$), on a logarithmic scale. For $r=1$ the
widths use the pre-registered upper edges, which the bias of Sec.~\ref{sec:res_23} raises by about $0.008$ in $g$, 10--30\%
of the width.}
\label{fig:window}
\end{figure*}

$\langle23\rangle$ passes at six of the seven points, mostly by a wide margin (Table~\ref{tab:s5a},
Fig.~\ref{fig:f23}). Its gain over $\langle2\rangle$ agrees with the series at every point, at sixth as at eighth
order (Table~\ref{tab:s5a8}). Its gain over $\langle3\rangle$ exceeds the sixth-order prediction at most points; for
decoupled layers the eighth order accounts for this, in the bilayer it does not (see below). The one failure is the
point with the smallest predicted margin, $\kappa=0.52$ at $r=1$, where $\langle23\rangle$ lies below $\langle2\rangle$
by $1.8$ standard errors against the required $2.12$, although the measured gain agrees with the prediction. With the
realized errors, passing all fourteen comparisons had a probability of only $0.58$ even if the theory is exact
(Appendix~\ref{app:prereg}). A follow-up at this point with 24 new chains per structure, decided after the result and
reported separately, resolves $\langle23\rangle$ ($t=-4.6$ and $-7.5$; Table~S3 of the Supplemental Material~\cite{supp}); the pooled values of
Fig.~\ref{fig:window}, chosen after the fact, carry no valid $p$-value.

The lower edge of the window agrees with the series at every point (Fig.~\ref{fig:window}). The upper edge lies above
the sixth-order value at six of the seven points, by more than $2\sigma$ at four, including both points at
$\kappa=0.53$, where the pre-registration anticipated agreement within $2\sigma$. For decoupled layers the eighth order removes most of this excess; in the bilayer the runs described next trace it to
the pre-registered $20\times30$ runs.
At $\kappa=0.5625$ the upper edge is not the end of the
phase: $\langle233\rangle$ drops below $\langle23\rangle$ first (Sec.~\ref{sec:res_beyond}). The
ratio of the window widths for $r=0$ and $r=1$ agrees with the eighth-order series [Fig.~\ref{fig:window}(c)]; the
second layer narrows the $\langle23\rangle$ window several times more strongly than the $\langle3\rangle$ wedge at
fixed $g$.

Further runs, decided after the test, check these results; the Supplemental Material~\cite{supp} (Sec.~S3 and
Table~S3) gives them in full, with the error budget. On a common $20\times60$ lattice, for both $r$ and with two
independent runs in the bilayer, both edges and both gains agree with the eighth-order series within $2\sigma$, and at
$\kappa=0.52$, $r=1$ the narrowest pre-registered margins become clear ones: $\langle23\rangle$ lies
$1.6$--$1.9\times10^{-4}$ per spin below $\langle3\rangle$ (five to nine standard errors) and $1.6\times10^{-4}$ below
$\langle2\rangle$ ($6.5$ standard errors). The one discrepancy is in the pre-registered bilayer comparisons with
$\langle3\rangle$ on $20\times30$ lattices at the standard number of sweeps, which favor $\langle23\rangle$ by about
$10^{-4}$ per spin too much; with four times the sweeps they agree with the common lattice and with the series, and we
think those runs were not fully equilibrated. Runs repeated with identical settings and new seeds agree within their
statistical errors, $\langle23\rangle$ remains below both neighbors at $T=0.10\,J_0$, and $40\times40$ bilayers
reproduce the lower edge. The statistical errors, $0.3$--$1.2\times10^{-4}$ per spin, dominate the error budget except
for $r=0$ at $\kappa\ge0.545$, where the truncation of the series and thermal excitations of the rows next to the walls
reach several $10^{-4}$ and the comparison of the upper edge with the series is qualitative. $\langle23\rangle$ lies
below both neighbors in every run and on every lattice at all seven points, the energies at $g_0$ agree with the series
[Fig.~\ref{fig:validation}(b)], and no chain left its structure before $g_c$.

Structures that the theory places just above $\langle23\rangle$ at $g_c$, namely $\langle223\rangle$, $\langle233\rangle$ and
$\langle2333\rangle$ (on $20\times28$, $20\times32$ and $20\times22$), lie above it at the four points where we followed
them, by amounts that agree with the series within about $2\sigma$ once the eighth order is included.

\subsection{Beyond the perturbative regime}
\label{sec:res_beyond}

All fields analysed here lie inside the ordered region: along the anneals on $12\times12$ lattices order sets in at
$g\approx2.2$--$2.4$ for $r=0$ and $3.1$--$3.4$ for $r=1$, above every crossing of Table~\ref{tab:crossings}. We do not
locate the paramagnetic boundary itself (Sec.~\ref{sec:validation}).

Beyond $\lambda=0.5$ the crossings move increasingly above the series (Table~\ref{tab:crossings}), and the eighth
order does not close the gap. The bilayer at $\kappa=0.75$ shows what this means for the phase diagram. The
leading-order wedge expanded about $\kappa=1/2$, Eq.~\eqref{eq:wedge}, places the $\langle3\rangle|\langle2\rangle$ crossing at
$g=3.46$, where order
sets in, and would leave no field range for $\langle3\rangle$; in the simulations $\langle3\rangle$ is the lowest of the
structures compared from $g^\star=2.69$ up to the onset of order.

Above the $\langle23\rangle$ window the simulations point to further structures [Fig.~\ref{fig:f23}(c)].
At $\kappa=0.5625$ and $r=0$, $\langle233\rangle$ drops below $\langle23\rangle$ before $\langle3\rangle$ does
(Table~\ref{tab:s5a}), and $\langle2333\rangle$ is lower still from $g\approx1.3$, so with increasing field the lowest of
the structures we followed is $\langle2\rangle$, $\langle23\rangle$, $\langle233\rangle$, $\langle2333\rangle$. These
fields lie at $\lambda=0.53$--$0.71$, beyond the range of the series. The sixth-order series places the
$\langle23\rangle|\langle233\rangle$ crossing close to the measured one, but the eighth order moves it above $g=1.6$, so
this agreement is fortuitous; at $g=1.2$ the series fails by about $10^{-3}$ per spin.
Closer to the multiphase point the effect fades: at $\kappa=0.545$ and $0.53$ the range in which $\langle233\rangle$ is
lowest is narrower than its uncertainty, and in the bilayer at $\kappa=0.53$ $\langle3\rangle$ overtakes
$\langle23\rangle$ before $\langle233\rangle$ does.

We report these structures as an indication only. Only five structures were compared, so a longer structure
could be lower still. The gaps at $g=1.2$ are several times the bias of about $10^{-4}$ found in some runs at the standard
sweeps (Sec.~\ref{sec:res_23}), but the $\langle233\rangle|\langle2333\rangle$ crossing is not robust to a shift of that
size, which would move it by about $0.08$. The walls may also soften at these fields (Sec.~\ref{sec:pt46}), which
the template branches would not capture. And the test was exploratory (T8 in Appendix~\ref{app:prereg}).
We therefore regard $\langle233\rangle$ and $\langle2333\rangle$ as the lowest of the structures we followed, not as
established phases.

Near the paramagnet the order may become incommensurate or floating, as in the classical two-dimensional
model~\cite{villain1981two,hu2021resolving}. This regime lies beyond our lattices and templates
(Sec.~\ref{sec:conclusion}).

\section{Discussion and conclusion}
\label{sec:conclusion}

For spin $1/2$ in a transverse field on the square lattice, quantum fluctuations select
$\langle3\rangle$ at the multiphase point at order $g^2$. The mechanism is local: spins next to a straight domain wall
have the smallest molecular field and gain the most energy from virtual flips, and domains of three sites carry the
largest number of such spins per site among domains that avoid the two-site penalty. The interlayer coupling raises
every local field by the same amount and weakens the selection. At the $\langle3\rangle|\langle2\rangle$ boundary,
fourth, sixth and eighth order then select $\langle23\rangle$, $\langle223\rangle$ and $\langle2223\rangle$. The
simulations confirm the two structures that are wide enough to resolve, for $\lambda\lesssim0.5$: the slope of the
$\langle3\rangle$ wedge to about $1\%$, and, in post-hoc runs on a common lattice at four points, both edges of the
$\langle23\rangle$ window within their statistical errors. $\langle23\rangle$ is the lowest structure at all seven test
points in every run.

The order of appearance is that of the Fisher--Selke analysis of the classical three-dimensional
model~\cite{fisher1980infinitely,fisher1981low,selke1988annni}, with $g^2$ in place of the low-temperature expansion
parameter (Sec.~\ref{sec:pt46}); Ref.~\cite{yeomans1995splitting} compares thermal and quantum splitting of the
multiphase point in the large-spin model. Unlike the large-spin Heisenberg analog, where spin deviations propagate
between walls~\cite{harris1995quantum,harris1995quantumb}, and the transverse chain, where mobile walls produce a floating
phase already at first order in $g$~\cite{chandra2007floating}, the walls here are rigid: a transverse field creates
only local flips, so the order $g^{2n}$ couples at most $n$ consecutive domains. The same rigidity limits the series,
because the rows next to the walls are themselves transverse-field chains or ladders, which become critical at $g=1$
($r=0$) and $g\approx1.83$ ($r=1$). The interlayer coupling suppresses the phases that appear at higher order more
strongly than $\langle3\rangle$. Between $r=0$ and $r=1$ the $\langle3\rangle$ wedge narrows by $2.4$ and the
$\langle23\rangle$ window by $6.7$ at fixed $g$; most of this reflects the larger local fields, and at fixed $\lambda$ the
factors are $1.07$, $1.32$, $1.58$ and $1.86$ for $\langle3\rangle$, $\langle23\rangle$, $\langle223\rangle$ and
$\langle2223\rangle$.

We establish the first four members of the sequence, and the work points to several next steps.
(i) Higher orders of the series would show whether further phases $\langle2^k3\rangle$ follow on the $\langle2\rangle$
side, as in the classical problem, and would decide the $\langle23\rangle|\langle223\rangle$ edge, which is still degenerate
at order $g^8$. (ii) Above the $\langle23\rangle$ window, beyond the range of the series, the simulations find
$\langle233\rangle$ and then $\langle2333\rangle$ below $\langle23\rangle$ and $\langle3\rangle$ at $\kappa=0.5625$. The
template method compares only the structures we follow, so a systematic search over longer sequences would identify the
stable structure there. (iii) On $20\times30$ bilayers at the standard number of sweeps the gain of $\langle23\rangle$ over
$\langle3\rangle$ comes out about $10^{-4}J_0$ per spin larger than with four times the sweeps or on $20\times60$
lattices; understanding why these runs equilibrate more slowly would sharpen the error model of the template method. (iv) Near the paramagnetic boundary the order may become floating, as in the
classical two-dimensional model~\cite{villain1981two,hu2021resolving}. Locating this boundary calls for larger lattices
and a finite-size analysis, which our test on the transverse-field Ising model shows to be demanding at this
temperature.

Annealing ten times more slowly gives the same result on $12\times12$ lattices, so the walls do not relax at the rates
we can afford; we used one update scheme throughout, and other schemes may fare differently. Annealing from the paramagnet freezes at the
wavevector at which order first develops, the allowed wavevector nearest the instability of the paramagnet, and
the rigid walls then prevent any reorganization at lower field. On $12\times12$ lattices this selects $\langle3\rangle$
at every $\kappa$ we studied, including where the ferromagnet or $\langle2\rangle$ is stable. On $12\times60$ lattices at
$\kappa=0.625$ and $0.75$ it selects sequences of two- and three-site domains with the corresponding wall densities.
In the bilayer at $\kappa=0.75$, where the leading-order wedge, expanded about $\kappa=1/2$, would have placed the $\langle3\rangle|\langle2\rangle$
crossing at the ordering field, no anneal reaches $\langle2\rangle$, although $\langle2\rangle$ is lower than $\langle3\rangle$
below $g^\star=2.69$. Similar obstacles may affect annealing protocols on programmable quantum devices that realize
frustrated transverse-field Ising models~\cite{king2018observation}.

\begin{acknowledgments}
We thank Can Tho University for its support. Claude Opus 5.5 (Anthropic) was used throughout the preparation of this
work: to search and survey the literature, to write and run the simulation and analysis code, to check the perturbative
coefficients independently, and to check the numerical claims and the characterizations of prior work against the data
and sources.
\end{acknowledgments}

\section*{Data availability}
The code, the result files and the scripts that reproduce every number and figure in this paper are available at
\url{https://github.com/Identity-Operator/transverse-field-annni-bilayer} and archived on Zenodo at
\href{https://doi.org/10.5281/zenodo.23133828}{doi:10.5281/zenodo.23133828}~\cite{truong2026zenodo}.

\appendix

\section{Minimum of the domain-sequence energy}
\label{app:pt}

Write the per-domain cost as $c(n)=(2-4\kappa)-\tfrac{g^2}{2}F(n)$, so that
$\Delta e=\sum_jc(n_j)/\sum_jn_j$ by Eq.~\eqref{eq:de}. For positive $n_j$,
$\sum_jc(n_j)/\sum_jn_j\ge\min_jc(n_j)/n_j$, with equality when every domain has the same ratio
$c(n_j)/n_j$. If $c(n)>0$ for all $n$, the minimum is the ferromagnet, $\Delta e=0$. Otherwise it is
reached by the periodic structure that minimizes $c(n)/n$. From Eq.~\eqref{eq:F},
\begin{equation}
F(3)-F(n\ge4)=\frac{2}{(3+r)(4+r)(5+r)}>0 .
\end{equation}
With $F$ constant for $n\ge4$, $\langle n\rangle$ with $n\ge4$ can beat both the ferromagnet and
$\langle3\rangle$ only if $F(n)>F(3)$, which never holds. The remaining comparisons, FM against
$\langle3\rangle$ ($c(3)<0$) and $\langle3\rangle$ against $\langle2\rangle$
($c(3)/3<c(2)/2$), give Eq.~\eqref{eq:wedge}. At $\kappa_{3|2}$ one has $c(2)/2=c(3)/3$, and any
sequence of two- and three-site domains has the same $\Delta e$.

\paragraph*{Higher orders.}
At order $g^4$ on the line $k=F(3)/4$ the same ratio structure appears, with the domain costs replaced by
costs of adjacent domain pairs. Let $P(D,D',B)=-4B(D+D')/[(DD')^2(D+D'-4B)]$ be the pair term of
Eq.~\eqref{eq:eps4}. Let $D_1,\dots,D_n$ be the flip costs along a domain (Table~\ref{tab:h}), and let
$\gamma_i=s_i(s_{i+2}+s_{i-2})$, which takes the values $-2$, $0$ and $2$. Let $X_n$ and $Y_n$ be the flip costs of
the first and second site from a wall of a domain of length $n$. Then, for a period of $L$ sites,
\begin{align}
W_4L&=\sum_j\big[a(n_j)+b(n_j,n_{j+1})\big],\nonumber\\
a(n)&=\sum_{i=1}^{n}\Big[-\frac{2k\gamma_i}{D_i^2}+\frac1{D_i^3}\Big]\nonumber\\
&\quad+\sum_{i=1}^{n}\Big[P(D_i,D_i,1)+\tfrac12P(D_i,D_i,r)\Big]\nonumber\\
&\quad+\sum_{i=1}^{n-1}P(D_i,D_{i+1},1)+\sum_{i=1}^{n-2}P(D_i,D_{i+2},-\tfrac12),\nonumber\\
b(n,n')&=P(X_n,X_{n'},-1)+P(Y_n,X_{n'},\tfrac12)\nonumber\\
&\quad+P(X_n,Y_{n'},\tfrac12).
\label{eq:ab}
\end{align}
With $C(n,n')=k_1(n-4)+a(n)+b(n,n')$, the fourth-order energy per spin is $\sum_jC(n_j,n_{j+1})/\sum_jn_j$.
A periodic sequence is a closed walk on the directed graph with vertices $\{2,3\}$ and all four edges, where
edge $n\to n'$ has cost $C(n,n')$ and length $n$. Every closed walk splits into elementary cycles. A ratio of
sums with positive denominators is at least the smallest ratio of its parts, with equality only if all parts
share that ratio. The elementary cycles are $2\to2$, $3\to3$ and $2\to3\to2$, that is $\langle2\rangle$,
$\langle3\rangle$ and $\langle23\rangle$, so one of them minimizes the energy for every $k_1$.

Now count the adjacent pairs $N_{22}$, $N_{33}$ and $N_{23}=N_{32}$ of a cyclic sequence. The period is
$L=2N_{22}+3N_{33}+5N_{23}$, and the total cost is $N_{22}C(2,2)+N_{33}C(3,3)+N_{23}[C(2,3)+C(3,2)]$. The point
$(\rho,W_4)$ of the sequence is therefore the average of the points of $\langle2\rangle$, $\langle3\rangle$ and
$\langle23\rangle$, weighted by the site fractions $2N_{22}/L$, $3N_{33}/L$ and $5N_{23}/L$. Sequences with
$N_{33}=0$ lie on the segment from $\langle2\rangle$ to $\langle23\rangle$, and sequences with $N_{22}=0$ lie on
the segment from $\langle23\rangle$ to $\langle3\rangle$. All other sequences lie strictly above the hull,
because $\langle23\rangle$ lies strictly below the chord from $\langle3\rangle$ to $\langle2\rangle$.

The argument rests on a locality property. Domains have at least two sites and every coupling spans at most two
sites, so a connected cluster of $n$ spins touches at most $n$ consecutive domains, and at $\kappa=1/2$ the field of
a spin depends only on its own domain (Table~\ref{tab:h}). The energy at order $g^{2n}$ is therefore a sum of terms for
$n$ consecutive domains. Non-periodic sequences are covered as limits of long periodic segments, whose energy obeys the
same bound. At order $g^6$ the terms of Eq.~\eqref{eq:W} involve at most three consecutive domains. The same argument,
applied to the graph whose vertices are pairs of consecutive domains, leaves $\langle2\rangle$,
$\langle23\rangle$ and $\langle223\rangle$ as the only candidates on the $\langle23\rangle|\langle2\rangle$
line, and $\langle3\rangle$, $\langle23\rangle$ and $\langle233\rangle$ on the $\langle3\rangle|\langle23\rangle$
line.

At order $g^8$ the terms involve at most four consecutive domains, and the argument applies to the graph whose
vertices are triples of consecutive domains. On the $\langle223\rangle|\langle2\rangle$ line the sequences degenerate
at order $g^6$ have isolated three-site domains separated by at least two two-site domains, and their elementary
cycles are $\langle2\rangle$, $\langle223\rangle$ and $\langle2223\rangle$; longer sequences such as $\langle22223\rangle$ or $\langle2232223\rangle$ decompose into
these and lie on the segments between them. The explicit evaluation places $\langle2223\rangle$ below the chord, so
it is the one structure that enters this line at order $g^8$ (Sec.~\ref{sec:pt46}). On the
$\langle23\rangle|\langle223\rangle$ line the sequences degenerate at order $g^6$ are built from the units $23$ and $223$.
Their triples of consecutive domains are $(2,3,2)$, $(3,2,3)$, $(3,2,2)$ and $(2,2,3)$, and the only elementary cycles
are $\langle23\rangle$ and $\langle223\rangle$, both through $(2,3,2)$. Every such sequence therefore lies on the segment
between them at order $g^8$, and no structure enters this line at this order.

\section{SSE implementation and validation}
\label{app:sse}

Bonds are grouped into classes with a common $|J_b|$ (bonds along $x$, bonds along $y$,
next-nearest-neighbor bonds along $y$, rungs). An insertion first chooses a class with probability
proportional to its total weight and then a bond uniformly within it, and is rejected when the
chosen bond is unsatisfied. The string length grows as $M=n+n/2+64$ during thermalization. Each
chain starts with 200 diagonal-only sweeps at its first field. Without them, the first cluster
update on a nearly empty string would flip every operator-free spin with probability $1/2$ and
randomize the initial structure. Diagonal updates cannot change the spins, so this step is part
of the thermalization. Every later update satisfies detailed balance with respect to the SSE weight, so a
chain samples that weight restricted to the basin of its initial structure for as long as it stays in it; chains
are cut when their dominant wavevector changes. The diagonal-only sweeps only place a chain in that basin.

The protocol and the full comparison of the exact-diagonalization benchmark are given in the Supplemental
Material~\cite{supp} (Table~S1).

\begin{figure*}
\includegraphics[width=\textwidth]{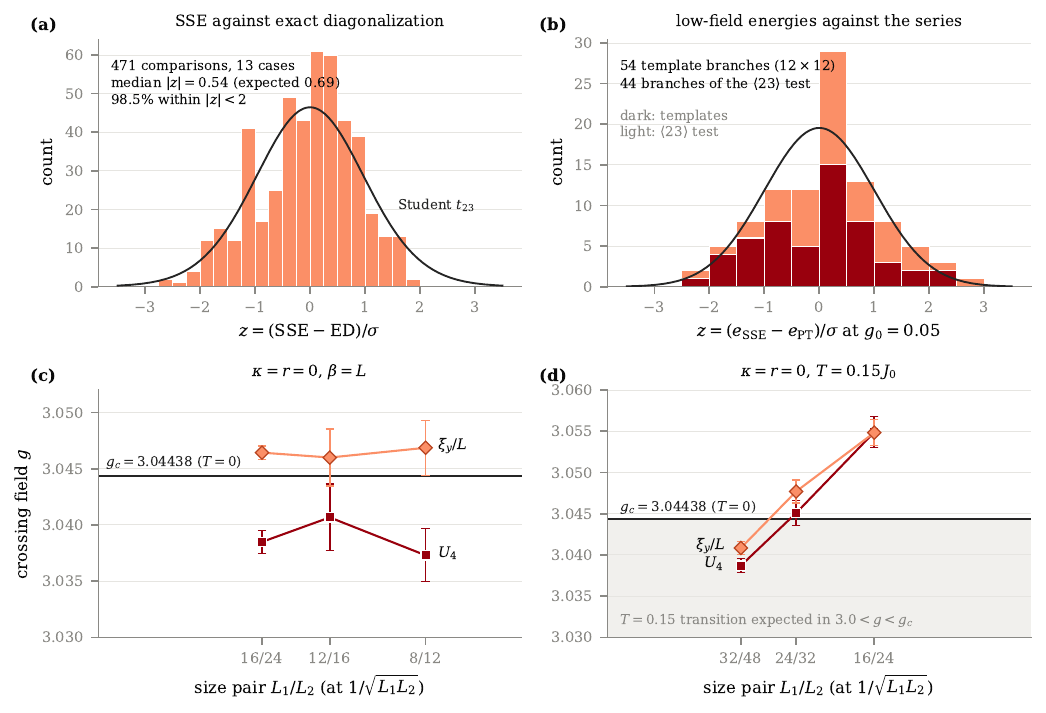}
\caption{Validation. (a) Pulls $z$ of all 471 comparisons of SSE with exact diagonalization
(Table~S1 of the Supplemental Material~\cite{supp}), against the Student-$t$ distribution with 23 degrees of freedom expected for averages of
24 chains. (b) Pulls of the measured energy at $g_0=0.05$ against the perturbation series, for the 54 template
branches of Sec.~\ref{sec:res_branches} (dark) and the 44 branches of the $\langle23\rangle$ test (light), with
the standard normal density. (c), (d) Two-dimensional transverse-field Ising model: crossing fields of the Binder
cumulant $U_4$ and of $\xi_y/L$ for pairs of sizes, (c) with $\beta=L$ and (d) at $T=0.15\,J_0$. The horizontal
line is the quantum critical point $g_c=3.04438(2)$~\cite{blote2002cluster}. At $T=0.15\,J_0$ the transition
lies between $g=3.0$~\cite{hesselmann2016thermal} and $g_c$ (shaded).}
\label{fig:validation}
\end{figure*}

Figure~\ref{fig:validation} summarizes the checks of Sec.~\ref{sec:validation}.

\section{Pre-registered test of the $\langle23\rangle$ phase}
\label{app:prereg}

The test of Sec.~\ref{sec:res_23} followed a written plan. Its design, analysis and pass criteria were
fixed before the production runs, after an independent review of the design; the full plan, the outcomes of all its
criteria and the change log are given in the Supplemental Material~\cite{supp} (Sec.~S2 and Table~S2). At each of seven
points $(\kappa,r)$ with $\lambda\le0.5$, the pairs $\{\langle23\rangle,\langle2\rangle\}$ ran on a $20\times20$ bilayer
and $\{\langle23\rangle,\langle3\rangle\}$ on a $20\times30$ bilayer, with 12 chains per structure and $2000+3000$ sweeps
per field at $T=0.15\,J_0$. The criteria used in the main text are (T1) $F_{\langle23\rangle}<F_{\langle2\rangle}$ and
$F_{\langle23\rangle}<F_{\langle3\rangle}$ at $g_c$, each with one-sided Welch $p<0.0228$ and no chain cut before $g_c$,
required at all seven points; (T6) consistency of $\langle23\rangle$ on the two lattices, $\max|z|<3$; and (T8) an
exploratory search for further structures above the window. The stop rule, no claim for $\langle23\rangle$ if a
competitor lay more than $3\sigma$ below it at $g_c$, was never triggered. At $\kappa=0.53$, $r=1$, T6 passed its
criterion, but with $\chi^2=3.45$ per point and 14 of 25 fields beyond $2\sigma$; the original $20\times30$ run for
$\langle23\rangle$ at this point is the one that the repeat with four times the sweeps did not reproduce
(Sec.~\ref{sec:res_23}).

\paragraph*{Power, computed after the test.} With the planned errors the probability of passing all fourteen
comparisons, if the theory is exact, was between $0.5$ and $0.7$; with the realized errors it was $0.58$. The headline
criterion was therefore underpowered, and it was not met.

The follow-up at $\kappa=0.52$, $r=1$ and the common-lattice runs were decided after the result of T1 was known, and
are reported separately.

\clearpage
\bibliography{refs}

\end{document}


\title{Supplemental Material for ``Onset of a quantum Fisher--Selke sequence in the transverse-field ANNNI model and its
bilayer''}
\maketitle

\section{Exact-diagonalization benchmark}

Table~\ref{tab:ed} compares the SSE code with exact diagonalization at finite temperature on four 12-spin periodic
clusters built by the production lattice code (Sec.~IV~C of the main text). The benchmark anneals every case over 16 fields from $g=3$ to the target,
with 2000 sweeps per field, followed by 5000 thermalization and 40\,000 measurement sweeps at the
target, in 24 independent chains.

\begin{table*}
\caption{SSE against exact diagonalization on 12-spin periodic clusters. $e$ is the energy per spin
and $m_x$ the transverse magnetization, both in units of $J_0$; the SSE values are means of 24
chains with standard errors in parentheses. The last column is the largest $|z|$ over all
observables of the case (31--39 per case).}
\label{tab:ed}
\begin{ruledtabular}
\begin{tabular}{llcrrrrlllll}
Cluster & & $T/J_0$ & $J_y/J_0$ & $\kappa$ & $\rp$ & $g$ & $e$ (ED) & $e$ (SSE) & $m_x$ (ED) & $m_x$ (SSE) & max$|z|$ \\
\hline
$2\times3$ bilayer & & 0.15 & 1 & 0 & 1 & 1.0 & $-2.60042$ & $-2.60075(20)$ & 0.20169 & 0.20176(14) & 1.65 \\
$2\times3$ bilayer & & 0.15 & 1 & 0.75 & $-1$ & 0.5 & $-1.78588$ & $-1.78603(17)$ & 0.14416 & 0.14394(23) & 2.01 \\
$2\times3$ bilayer & & 0.50 & 1 & 0.5 & 1 & 2.0 & $-2.52091$ & $-2.52076(68)$ & 0.54569 & 0.54629(57) & 2.62 \\
$2\times6$ layer & & 0.15 & 1 & 0 & 0 & 2.0 & $-2.52595$ & $-2.52599(23)$ & 0.56144 & 0.56135(18) & 0.68 \\
$2\times6$ layer & & 0.15 & 1 & 0.75 & 0 & 1.0 & $-1.83990$ & $-1.84065(41)$ & 0.46098 & 0.46162(69) & 1.83 \\
$2\times6$ layer & & 0.15 & 1 & 0.5625 & 0 & 1.5 & $-2.07157$ & $-2.07171(41)$ & 0.64328 & 0.64283(47) & 1.86 \\
$2\times6$ layer & & 0.15 & $-1$ & 0.25 & 0 & 1.0 & $-1.89600$ & $-1.89616(20)$ & 0.29902 & 0.29881(20) & 1.82 \\
$2\times6$ layer & & 0.50 & 1 & 1.0 & 0 & 0.5 & $-1.71530$ & $-1.71590(69)$ & 0.18847 & 0.18876(157) & 2.00 \\
$1\times6$ ladder & & 0.15 & 1 & 0.75 & 1 & 1.0 & $-1.46156$ & $-1.46128(36)$ & 0.61156 & 0.61140(57) & 1.80 \\
$1\times6$ ladder & & 0.15 & 1 & 1.3 & $-1$ & 0.3 & $-1.30261$ & $-1.30361(86)$ & 0.22459 & 0.22781(527) & 2.50 \\
$1\times6$ ladder & & 0.15 & 1 & 0.5 & 1 & 3.0 & $-3.13579$ & $-3.13617(44)$ & 0.95525 & 0.95528(12) & 1.68 \\
$3\times4$ layer & & 0.15 & 1 & 0.375 & 0 & 2.5 & $-2.74079$ & $-2.74034(26)$ & 0.88540 & 0.88520(14) & 1.68 \\
$3\times4$ layer & & 0.15 & 1 & 1.0 & 0 & 1.0 & $-2.12664$ & $-2.12642(20)$ & 0.25671 & 0.25629(20) & 2.08 \\
\end{tabular}
\end{ruledtabular}
\end{table*}

\section{Pre-registered test of the $\langle23\rangle$ phase}

The test of Sec.~V~B of the main text followed a written plan. Its design, analysis and pass criteria were fixed before
the production runs, after an independent review of the design.

\paragraph*{Design.} Seven points $(\kappa,r)$ with $\lambda\le0.5$, chosen so that the predicted margin
of $\langle23\rangle$ at $g_c$ exceeds the expected statistical error. At each point the pairs
$\{\langle23\rangle,\langle2\rangle\}$ ran on a $20\times20$ bilayer and $\{\langle23\rangle,\langle3\rangle\}$ on
a $20\times30$ bilayer, with 12 chains per structure, $2000+3000$ sweeps per field at $T=0.15\,J_0$, and
the ascending template protocol of Sec.~IV~B of the main text. The field grid had steps of $0.1$ below the
predicted window, steps of $0.02$ (or $0.01$) across it with a margin of at least $0.06$, and $g_c$ as an
exact grid point. Competitors ($\langle223\rangle$, $\langle233\rangle$, $\langle2333\rangle$) ran at four of
the points, and one point was repeated at $T=0.10\,J_0$.

\paragraph*{Analysis.} Free energies as in Sec.~IV~B of the main text. Uncertainties from the jackknife
over chains; differences tested with Welch's $t$ and its Satterthwaite degrees of freedom; crossings
from a cubic-spline root of the free-energy difference, with linear interpolation in $g^2$ as a cross
check.

\paragraph*{Criteria.}
(T1)~Existence at each point: $F_{\langle23\rangle}<F_{\langle2\rangle}$ and $F_{\langle23\rangle}<F_{\langle3\rangle}$
at $g_c$, each with one-sided $p<0.0228$, and no chain cut before $g_c$. The headline required a pass at
all seven points, and a failure at $\lambda\le0.4$ was to be reported and investigated.
(T2)~Window edges compared with the exact-$\kappa$ $O(g^4)$ and $O(g^6)$ values (not pass/fail).
(T3, T4)~Width of the window and the ratio of widths for $r=0$ and $1$.
(T5)~Competitors at $g_c$ against their $O(g^6)$ values, with a stop rule: no claim for
$\langle23\rangle$ if a competitor lay below it at $g_c$ by more than $3\sigma$.
(T6)~Consistency of $\langle23\rangle$ on the two lattices ($\max|z|<3$) and of $\langle3\rangle$ with the
$24\times24$ branches.
(T7)~Window edges at $T=0.10$ and $0.15$ agree within $2\sigma$.
(T8)~An exploratory search for further structures above the window.

Table~\ref{tab:prereg} lists the outcomes. The stop rule of T5 was never triggered. At $\kappa=0.53$, $r=1$, T6
passed its criterion, but with $\chi^2=3.45$ per point and 14 of 25 fields beyond $2\sigma$; the original $20\times30$ run
for $\langle23\rangle$ at this point is the one that the repeat with four times the sweeps did not reproduce
(Sec.~V~B of the main text).

\begin{table}
\caption{Outcomes of the pre-registered criteria. T2: pulls of the lower and upper edges against $O(g^6)$. T5: largest
$|{\rm pull}|$ of a competitor against $O(g^6)$. T6: $\max|z|$ of $\langle23\rangle$ on $20\times20$ against
$20\times30$, with $\chi^2$ per point. T7: pulls of the edges at $T=0.10\,J_0$ against $T=0.15\,J_0$. T8: fields at which
$\langle233\rangle$ lies $2\sigma$ below $\langle23\rangle$ and $\langle3\rangle$. T3 and T4 are in
Table~VI of the main text and Fig.~5(c) of the main text.}
\label{tab:prereg}
\footnotesize
\begin{ruledtabular}
\begin{tabular}{llllllll}
$\kappa$ & $r$ & T1 & T2 & T5 & T6 & T7 & T8 \\
\hline
0.52   & 0 & pass & $-1.8$, $-1.1$  & ---   & 1.15 (0.57) & --- & --- \\
0.53   & 0 & pass & $+0.1$, $+2.7$  & 1.4   & 1.35 (0.59) & $-0.1$, $-1.7$ & none \\
0.545  & 0 & pass & $-0.7$, $+5.1$  & 1.2   & 1.85 (0.55) & --- & none \\
0.5625 & 0 & pass & $-0.9$, $+11.4$ & 3.3   & 1.04 (0.34) & --- & 1.10--1.30 \\
0.52   & 1 & fail & $+1.4$, $+1.2$  & ---   & 1.38 (1.14) & --- & --- \\
0.53   & 1 & pass & $-0.5$, $+2.9$  & 1.5   & 2.19 (3.45) & --- & none \\
0.545  & 1 & pass & $+0.4$, $+1.4$  & ---   & 1.15 (0.34) & --- & --- \\
\end{tabular}
\end{ruledtabular}
\end{table}

\paragraph*{Changes.} Before the runs, the review moved $\langle3\rangle$ from a $12\times12$ to the
$20\times30$ lattice, added $g_c$ to the grids, specified the quadrature, and replaced a test on
$\min(F_{\langle2\rangle},F_{\langle3\rangle})$ by the two separate comparisons of T1. After the first
output existed, eight ambiguities in the analysis were resolved in writing without changing the text of
the plan, and an error that affected only chains cut at a change of structure was corrected. It had no
effect on any chain used in T1. The follow-up at $\kappa=0.52$, $r=1$ was decided after the result of
T1 at that point was known, and is reported separately.

\section{Additional runs and error budget}

\subsection{Common lattice, replicates and sweeps}
\label{sec:s3runs}

After the pre-registered test (Sec.~V~B of the main text) we ran the following checks; Table~\ref{tab:lattices}
collects the results. For decoupled layers the common lattice at $\kappa=0.53$ places the upper edge $1.7\sigma$ above
the eighth-order value and $2.9\sigma$ above the sixth. In the bilayer the pre-registered upper edges lie above the
series, and the paired differences of $e_b-f_b$ between $\langle23\rangle$ and $\langle3\rangle$ are consistent with
zero, so the integration does not cause this excess.

To locate the bilayer excess we followed all three structures on one $20\times60$ lattice at both bilayer points, twice
with independent seeds, and repeated $\langle23\rangle$ and $\langle3\rangle$ with four times the sweeps on both lattices
at $\kappa=0.52$ (Table~\ref{tab:lattices}). For decoupled layers the common lattice agrees with the pre-registered
values and with the eighth-order series at both $\kappa$; at $\kappa=0.52$ it also turns the narrowest pre-registered
margin, $\langle23\rangle$ below $\langle3\rangle$, into a clear one. In the bilayer the two $20\times60$ runs agree with
each other, and together they place both edges and both gains within $2\sigma$ of the eighth-order series. The gain over
$\langle2\rangle$ is the same on every lattice. The gain over $\langle3\rangle$ is about $10^{-4}$ per spin larger on
$20\times30$ at the standard sweeps, in every set run with these settings: the original runs, the follow-up at
$\kappa=0.52$ and the repeat at $T=0.10\,J_0$. With four times the sweeps it falls to the $20\times60$ value at both
points, while on $20\times60$ it does not change. At $\kappa=0.52$ the four combinations of lattice and sweeps thus
place the shift in the one combination of $20\times30$ and standard sweeps (interaction $1.9\sigma$, simple effects
$2.1$--$2.5\sigma$), and we think these runs were not fully equilibrated. The original $20\times30$ run for
$\langle23\rangle$ at $\kappa=0.53$ had also been flagged by the consistency check T6 (Appendix~C of the main text). The
energies shift with the free energies, so the integration is not the cause, and the pair correlations along $y$ show no
wall defects. Runs repeated with identical settings and new seeds agree within their statistical errors (rms $t=1.06$
over ten comparisons). We therefore take the $20\times60$ and four-times-sweeps values as our estimates. They agree with
the eighth-order series within $1\sigma$ in the gain over $\langle3\rangle$ and the upper edge at both points; in the
bilayer the sixth and eighth orders differ by less than these errors. The pre-registered $\Delta F_3$ in the bilayer
carries a bias of about $-10^{-4}$ per spin, which raises the upper edge by about $0.008$ in $g$. $\langle23\rangle$ lies
below both neighbors in every run and on every lattice at all seven points. At $\kappa=0.52$, $r=1$ its margin over
$\langle3\rangle$ is $1.6$--$1.9\times10^{-4}$ per spin in the three unbiased sets, five to nine standard errors, and
its margin over $\langle2\rangle$ on the common lattice is $1.6\times10^{-4}$, $6.5$ standard errors.

At $T=0.10\,J_0$, where the thermal contributions estimated below are about ten times smaller, $\langle23\rangle$
remains below both neighbors at $\kappa=0.53$ for both $r$ (Table~\ref{tab:lattices}). For $r=0$ the lower edge and the
gain over $\langle2\rangle$ are unchanged, and the upper edge and the gain over $\langle3\rangle$ move by $1.7\sigma$ and
$1.5\sigma$; for $r=1$ both edges and both gains agree with those at $T=0.15\,J_0$ within $1\sigma$.

\begin{table}
\caption{Test points on different lattices and in different runs: window edges and the free-energy
differences $\Delta F_2$ and $\Delta F_3$ of $\langle23\rangle$ relative to $\langle2\rangle$ and $\langle3\rangle$ at $g_c$
(in $10^{-4}J_0$ per spin, negative where $\langle23\rangle$ is lower), with statistical errors. The
pre-registered runs use $20\times20$ for $\langle2\rangle$ and $20\times30$ for $\langle3\rangle$; the follow-up at
$\kappa=0.52$ was decided afterwards; ``$4\times$'' has four times the sweeps; $20\times60$ holds all three
structures, ``replicate'' repeats that run with new seeds, and ``pooled'' combines the chains of both; ``$T=0.10$'' repeats the
pre-registered runs at $T=0.10\,J_0$; $40\times40$ holds $\langle2\rangle$ and $\langle23\rangle$ only.}
\label{tab:lattices}
\footnotesize\setlength{\tabcolsep}{1.2pt}
\begin{ruledtabular}
\begin{tabular}{llllll}
$\kappa$, $r$ & run & $g_{2|23}$ & $g_{23|3}$ & $\Delta F_2$ & $\Delta F_3$ \\
\hline
0.52, 0 & pre-registered & 0.5885(21) & 0.6084(36) & $-3.32(60)$ & $-1.46(62)$ \\
     & $20\times60$ & 0.5917(13) & 0.6130(28) & $-2.34(38)$ & $-2.26(47)$ \\
     & $O(g^8)$ & 0.5922 & 0.6134 & $-2.21$ & $-2.32$ \\
0.53, 0 & pre-registered & 0.7050(19) & 0.7553(51) & $-4.93(76)$ & $-7.19(99)$ \\
     & $20\times60$ & 0.7038(11) & 0.7522(37) & $-5.41(42)$ & $-6.84(72)$ \\
     & $T=0.10$ & 0.7048(17) & 0.7451(33) & $-5.12(65)$ & $-5.36(66)$ \\
     & $O(g^8)$ & 0.7045 & 0.7460 & $-5.16$ & $-5.63$ \\
0.53, 1 & pre-registered & 1.1037(29) & 1.1604(27) & $-4.31(66)$ & $-4.90(34)$ \\
     & $20\times30$, $4\times$ & --- & 1.1550(29) & --- & $-4.23(37)$ \\
     & $20\times60$ & 1.1033(19) & 1.1483(39) & $-4.38(45)$ & $-3.38(51)$ \\
     & replicate & 1.1017(18) & 1.1552(32) & $-4.75(40)$ & $-4.25(43)$ \\
     & pooled & 1.1025(13) & 1.1517(26) & $-4.56(30)$ & $-3.82(34)$ \\
     & $40\times40$ & 1.1049(11) & --- & $-4.03(27)$ & --- \\
     & $T=0.10$ & 1.1069(17) & 1.1612(37) & $-3.56(39)$ & $-5.06(49)$ \\
     & $O(g^8)$ & 1.1046 & 1.1536 & $-4.08$ & $-4.09$ \\
0.52, 1 & pre-registered & 0.9297(30) & 0.9577(52) & $-0.99(54)$ & $-2.47(58)$ \\
     & follow-up & 0.9232(25) & 0.9620(36) & $-2.13(46)$ & $-2.90(39)$ \\
     & $20\times60$ & 0.9258(19) & 0.9477(40) & $-1.67(34)$ & $-1.38(44)$ \\
     & replicate & 0.9269(20) & 0.9527(38) & $-1.48(35)$ & $-1.90(41)$ \\
     & pooled & 0.9264(13) & 0.9502(28) & $-1.58(24)$ & $-1.64(30)$ \\
     & $20\times30$, $4\times$ & --- & 0.9526(23) & --- & $-1.91(25)$ \\
     & $20\times60$, $4\times$ & --- & 0.9524(20) & --- & $-1.87(22)$ \\
     & $O(g^8)$ & 0.9253 & 0.9515 & $-1.78$ & $-1.78$ \\
\end{tabular}
\end{ruledtabular}
\end{table}

\subsection{Error budget}

The error budget of the free-energy differences at the test points has five parts. The statistical errors are
$0.3$--$1.2\times10^{-4}$ per spin (Table~VI of the main text). The choice of quadrature (trapezoidal, Richardson or spline)
moves them by up to $0.4\times10^{-4}$. The truncation of the series, estimated by the last retained term, the
difference between the eighth- and sixth-order values of Tables~VI and VII of the main text, is at most
$0.5\times10^{-4}$ in the bilayer and $0.7\times10^{-4}$ for $r=0$ at $\kappa\le0.53$, but $2.6\times10^{-4}$ and
$7.6\times10^{-4}$ for $\Delta F_3$ at $\kappa=0.545$ and $0.5625$, $r=0$. Thermal excitations of the rows next to the
walls contribute at $T=0.15\,J_0$. For $r=0$ a free-chain estimate, which neglects the pinning of these rows by the walls
and therefore bounds the effect from above, gives up to $0.3\times10^{-4}$ at $\kappa=0.52$, $1.0\times10^{-4}$ at
$\kappa=0.53$ and several $10^{-4}$ at $\kappa\ge0.545$; it favors $\langle23\rangle$ over $\langle2\rangle$ and
$\langle3\rangle$ over $\langle23\rangle$, the opposite of the upper-edge excess. For $r=1$ the rows form ladders whose kink
energy at the test fields exceeds $1.2J_0$, and the estimate is below $10^{-5}$. Finally, the pre-registered $20\times30$ runs in the bilayer carry the
bias of about $-10^{-4}$ in $\Delta F_3$ discussed in Sec.~\ref{sec:s3runs}. For $r=0$ at $\kappa\ge0.545$ the comparison of $\Delta F_3$ and of the upper
edge with the series is therefore qualitative.